\documentclass[showpacs,twocolumn,superscriptaddress,nofootinbib]{revtex4-1} 

\usepackage[utf8]{inputenc}
\usepackage{hyperref}
\hypersetup{colorlinks=true,linkcolor=blue,citecolor=cyan}
\usepackage{graphicx}
\usepackage{xcolor, soul}
\sethlcolor{green}
\usepackage{dcolumn}
\usepackage{bm}
\usepackage{color}
\usepackage{enumitem}
\usepackage{mathrsfs}
\usepackage{amsmath}
\usepackage{amssymb}
\usepackage{cleveref}
\usepackage{orcidlink}

\crefname{equation}{Eqn.}{Eqns.}
\crefname{figure}{Fig.}{Figs.}
\crefname{section}{Sec.}{Sec.}
\crefname{table}{Table}{Tables}
\usepackage{physics}
\usepackage{multirow}
\usepackage{booktabs}
\usepackage{float}
\usepackage{afterpage}

\setstcolor{red}
\begin{document}

\title{\Large Observable thin accretion disk around a self-dual black hole \\in loop quantum gravity}

\author{Tursunali Xamidov}
\email{xamidovtursunali@gmail.com}
\affiliation{Institute of Fundamental and Applied Research, National Research University TIIAME, Kori Niyoziy 39, Tashkent 100000, Uzbekistan} 
\affiliation{Tashkent State Technical University, 100095 Tashkent, Uzbekistan}

\author{Javokhir Sharipov}
\email{javohirsh100@gmail.com}
\affiliation{Institute of Fundamental and Applied Research, National Research University TIIAME, Kori Niyoziy 39, Tashkent 100000, Uzbekistan}

\author{Sanjar Shaymatov}
\email{sanjar@astrin.uz}
\affiliation{Institute of Fundamental and Applied Research, National Research University TIIAME, Kori Niyoziy 39, Tashkent 100000, Uzbekistan}
\affiliation{Institute for Theoretical Physics \& Cosmology, Zhejiang University of Technology, Hangzhou 310023, China}
%\affiliation{University of Tashkent for Applied Sciences, Str. Gavhar 1, Tashkent 100149, Uzbekistan}
%\affiliation{Western Caspian University, Baku AZ1001, Azerbaijan}

\date{\today}
\begin{abstract}

In this paper, we study a self-dual black hole (BH) in Loop Quantum Gravity (LQG), analyzing both timelike and null geodesics. Using observational data from Mercury’s perihelion shift and the orbit of the S2 star around Sagittarius A$^{\star}$ (Sgr A$^{\star}$), we derive constraints on the polymeric function $P$. We further investigate photon trajectories near the self-dual BH under various scenarios to explore their observational relevance. Finally, we examine the properties of accretion disks around the self-dual BH in LQG, including their direct and secondary images, and study the redshift and the observed energy flux distribution across the accretion disk as measured by distant observers for different inclination angles. Our findings provide new insights into the physical nature and accretion properties of self-dual BHs in LQG and their possible observational consequences.

\end{abstract}

%\pacs{04.70.Bw, 04.20.Dw}
\maketitle

\section{Introduction}
\label{introduction}

Although general relativity (GR) stands as the most successful theory of gravity, it faces unresolved problems and significant challenges. Recent observations, including those involving VLBI \cite{Abbott16a,Abbott16b} and the Event Horizon Telescope (EHT) \cite{Akiyama19L1,Akiyama19L6}, have provided compelling evidence and opened a qualitatively new era in exploring the properties of astrophysical black holes (BHs). However, on the theoretical side, GR predicts spacetime singularities \cite{Penrose65a,Hawking-Penrose70} and lacks integration with quantum mechanics \cite{Adler10,Ng03MPLA}. Observationally, it cannot fully explain dark matter and dark energy. Singularities appear both in the early universe \cite{Borde03PRL,Borde94PRL} and inside black holes \cite{Hawking73}, where physics becomes inapplicable. These shortcomings motivate the exploration and testing of modified and quantum gravity theories as candidates for a more fundamental theory of gravity. 

To address these issues, Loop Quantum Gravity (LQG) provides a compelling resolution to the classical singularities of BHs and the universe. Within a minisuperspace framework, and using LQG polymerization procedure, a regular static metric known as the self-dual spacetime was addressed in Ref. \cite{Modesto10}. This geometry is free of curvature singularities, and its structure is determined by two intrinsic LQG parameters, such as the minimal area and the Barbero–Immirzi parameter. Furthermore, it exhibits a self-duality analogous to the T-duality \cite{Modesto09,Sahu15}, which motivates its designation as self-dual spacetime. In recent years, substantial progress has been made in the study of BHs in LQG with detailed discussions; see \cite{Ashtekar18PRL,Ashtekar18,Bojowald18,Alesci19PLB,Assanioussi20,Perez17,Barrau18,Rovelli18,Ashtekar20,Gan20}.

An important question is whether quantum gravity parameters or the LQG effects can produce observable signatures in current or forthcoming data, thereby enabling these effects on BH spacetimes to be directly tested or constrained. With this motivation, several observational and phenomenological implications of LQG BH spacetimes have already been investigated for the past decades in various situations \cite{Alesci14,Chen11, Barrau14, Dasgupta13,Sahu15,Cruz19,Moulin19a,Moulin19b,Cruz20,Santos21,Zhu20,Jiang24PDU,Uktamov25PDU}. It should be emphasized that the impact of LQG on the shadows of rotating BHs and gravitational lensing was investigated, with observational consequences analyzed associated with recent EHT observations of M87$^{\star}$ \cite{Liu20,Sahu15}. Furthermore, observational constraints on the LQG BH parameter (i.e., the polymeric parameter) were obtained from solar system tests of self-dual spacetimes \cite{Zhu20}. It is also noteworthy that numerous works have extensively examined observational implications and the phenomenology of various loop quantum BH models \cite{Liu21PRD,Daghigh21,Bouhmadi-Lopez20JCAP,Fu22,Brahma21PRL,shu2024qcbh,Konoplya25LQG,Liu24PLB,Liu24EQG,Du25JCAP,Wang25EPJC...85..302W}.  

It is also worth emphasizing that solutions of both GR and alternative theories of gravity can be probed through astrophysical processes. In particular, X-ray data \cite{Bambi12a,Bambi16b} and recent observations with accretion disk properties \cite{Abramowicz13,Fender04mnrs,Auchettl17ApJ} provide powerful avenues for probing the nature of gravity in strong-field regimes. 
Therefore, accretion disks around black holes are regarded as key probes of gravity and spacetime geometry in the strong field regime, as well as of high-energy processes occurring in their immediate vicinity~\cite{Abramowicz13}. Accretion disks form as matter spirals inward under gravity, losing angular momentum, and converting gravitational potential energy into radiation. Their study is essential in improving our understanding of high-energy astrophysical phenomena, including quasars~\cite{Czerny23ApSS,Morgan10ApJ} and X-ray binaries~\cite{Yao01AIP,Schultz05BOOK,Bu23BOOK}. 

Recent observational advances, particularly those enabled by the EHT, have delivered prominent insights into accretion flows around compact objects. In particular, Sagittarius A$^{\star}$ (Sgr A$^{\star}$) at the Galactic center serves as a unique laboratory for testing BH accretion physics, thus probing the strong-field regime of gravity and constraining BH models \cite{you2024,Nucita07,Ghez05,Ghez00,Imdiker23CQG,Vagnozzi23EHT,Afrin23ApJ}. In recent years, extensive studies of accretion disks within different gravity models have provided valuable insights into their behavior and the fundamental nature of gravitational interactions \cite{Boshkayev:2020kle,Gyulchev20,Gyulchev2021E,sym14091765,Shaymatov2023,Collodel_2021,Alloqulov24CPC,Boshkayev21PRD,Hu2023,Kurmanov24PDU,Cui24,Alloqulov24EPJP,Chen2025,Ziqiang25}. These studies collectively play a pivotal role in probing the strong field regime of gravity and constraining theoretical models of BHs. Given the significance of both accretion disks and Loop Quantum Gravity (LQG), considering a self-dual BH in LQG is particularly relevant, as it can modify test particle and null geodesics that can be used to model the accretion disk~\cite{Bambi17e,Chandrasekhar98}, and consequently affect observable quantities such as the innermost stable circular orbit (ISCO) and accretion disk properties. 
 
From an astrophysical perspective, LQG black holes differ fundamentally from their classical counterparts and provide a framework for probing potential quantum-gravitational signatures. Motivated by this, in this paper, we investigate accretion disks around self-dual black holes, including their images and the redshift and flux of emitted radiation as seen by distant observers. These analyses yield insights into the physical nature of the self-dual BH spacetime and their possible observational consequences. The radiation from accretion disks around LQG black holes would be particularly significant, as it may encode deviations from GR caused by quantum corrections arising in LQG. 

This paper is structured as follows. In Sec.~\ref{Sec:LQG}, we briefly introduce the metric of a self-dual BH in LQG and analyze the motion of massive particles around the BH and derive constraints on the quantum correction parameter $P$ (polymeric function) using observational data from Mercury’s perihelion shift and the S2 star around Sgr A$^{\star}$. This is followed by the study of the effect of $P$ on null geodesics with photon trajectories near self-dual BH in LQG in Sec.~\ref{Null}. In Sec.~\ref{sec:disk}, we focus on examining the properties of the accretion disk around the self-dual BH in LQG and its accretion images and discuss the redshift and observed radiation flux in the accretion disk relative to the observer, highlighting the main differences between the self-dual and Schwarzschild BHs. Finally, we end up with the conclusion in Sec.~\ref{Sec:conclusion}. We use the system of units $c=G=M=1$ throughout the paper.

\section{A self-dual BH spacetime in loop quantum gravity and time like particle geodesics}\label{Sec:LQG}

Here, we consider a self-dual BH spacetime, often referred to as the quantum-corrected Schwarzschild BH in LQG (see Ref.~\cite{Modesto10} for details). This spacetime arises through a polymer-like quantization procedure within LQG, which introduces a polymeric function $P$ associated with the polymeric parameter $\delta$ (or $\delta_b$). In this framework, the Schwarzschild interior is quantized by employing two parameters, $\delta_b$ and $\delta_c$, which serve to replace the corresponding canonical variables: 
\begin{eqnarray}
b \to \frac{\sin(\delta_b b)}{\delta_b} \;\mbox{and}\; c \to \frac{\sin(\delta_c c)}{\delta_c}\, .
\end{eqnarray}
It should be emphasized that $\delta_b$ and $\delta_c$ define the lengths of the paths along which holonomies are constructed, thereby setting the scales at which quantum effects become relevant. In the limit $(\delta_b,\delta_c) \to 0$, the classical Schwarzschild solution is recovered. Nevertheless, due to the absence of a complete theory of quantum gravity, no definitive prescription exists for fixing these parameters. As a result, multiple proposals have been suggested in the literature, each motivated by distinct physical or mathematical considerations. However, the self-dual spacetime considered in this work is constructed from a particular choice of fixing $\delta_b$ and $\delta_c$ as constants,  corresponding to the $\mu_0$-scheme in LQG~\cite{Ashtekar:2005qt, Modesto10}. This choice yields the effective quantum-corrected interior of the Schwarzschild black hole. By identifying the minimum area in the solution with the fundamental minimum area $A_{\text{min}}$ of the LQG, the two free parameters can be reduced. The remaining undetermined constant, $\delta_b$, represents the dimensionless polymeric parameter, which must be constrained by experiments or observations. 

Taken together, the effective self-dual BH spacetime metric in LQG, expressed in Schwarzschild coordinates, can be written as follows \cite{Modesto10}:  
\begin{eqnarray}
    {ds}^2=-A(r){dt}^2+\frac{{dr}^2}{B(r)}+C(r)({d\theta}^2+sin^2{\theta}{d\phi}^2)\, ,
    \label{metric}
\end{eqnarray}
with  
\begin{eqnarray}
    &A(r)&=\frac{(r-r_{+})(r-r_{-})(r+r_{*})^2}{r^4+a_{0}^2},\\
    &B(r)&=\frac{(r-r_{+})(r-r_{-})r^4}{(r+r_{*})^2(r^4+a_{0}^2)},\\
    &C(r)&=r^2+\frac{a_{0}^{2}}{r^2}\, ,
    \label{ABC}
\end{eqnarray}
where $r_{+}=2/(1+P)^2$ and $r_{-}=2P^2/(1+P)^2$ refer to the two BH horizons, respectively, and $r_{*}=2P/(1+P)^2$ with the polymeric function $P$, which is given by  
\begin{eqnarray}
    P\equiv\frac{\sqrt{1+\epsilon^2}-1}{\sqrt{1+\epsilon^2}+1}
    \label{P}\, .
\end{eqnarray}
Note that $\epsilon$ refers to the product of the Immirzi parameter $\gamma$ and the polymeric parameter $\delta$, i.e., $\epsilon=\delta \gamma \ll 1$ as a small quantity, while $a_0={A_{\text{min}}}/{8\pi}$
is given in terms of the LQG minimum area $A_{\text{min}} \approx 4\pi\gamma \sqrt{3}l^2_{PL}$. However, its contribution can be regarded as negligible as it is proportional to Planck length $l_{Pl}$. Consequently, $a_0=0$ can be set for further analysis.

\subsection{The geodesics of a test particle}

The Hamiltonian governing the motion of a particle in the vicinity of the self-dual BH in LQG is given by \cite{Misner73}:
\begin{equation}\label{hamiltonian}
H=\frac{1}{2}g^{\alpha\beta}p_\alpha p_\beta \ ,
\end{equation}
where $p^\alpha=m\,u^\alpha$ denotes the four-momentum with $u^\alpha = dx^\alpha/d\tau$ the four-velocity of the particle. In spherical coordinates, the indices $\alpha$ and $\beta$ correspond to the coordinate components $(t, r, \theta, \phi)$. The Hamiltonian for a massive particle, $H = -m^2/2$, while for a massless particle (e.g., a photon), $H = 0$. The corresponding Hamilton's equations are:
\begin{equation} \label{Hamx}
    \frac{dx^\alpha}{d\lambda} = \frac{\partial H}{\partial p_\alpha} \ ,
\end{equation}
\begin{equation} \label{Hamp}
    \frac{dp_\alpha}{d\lambda} = -\frac{\partial H}{\partial x^\alpha} \ ,
\end{equation}
with $\lambda$ denoting the affine parameter, related to the proper time $\tau$ via $\lambda = \tau / m$.

Since the spacetime metric \eqref{metric} is independent of the coordinates $t$ and $\phi$, the corresponding conserved quantities can be identified by the energy $p_t = -E$ and the angular momentum $p_\phi = L$ of the particle, respectively \cite{Misner73}. For that the Hamiltonian of the system yields:
\begin{align} \label{ex-hamiltonian}
    H = \frac{1}{2} \left( g^{rr} p_r^2 + g^{\theta\theta} p_\theta^2 + g^{tt} E^2 + g^{\phi\phi} L^2 \right) .
\end{align}

For a massive particle of rest mass $m$, we shall further restrict the motion to the equatorial plane (i.e., $\theta=\pi/2$). Subsequently, the radial equation of motion reads as
\begin{align} \label{radial}
   \left(\frac{dr}{d\tau}\right)^2 
   = -\frac{g^{tt}\mathcal{E}^2 + g^{\phi\phi}\mathcal{L}^2 + 1}{g_{rr}} \, ,
\end{align}
where $\mathcal{E}=E/m$ and $\mathcal{L}=L/m$ denote the specific energy and angular momentum.  
Meanwhile, the azimuthal equation of motion is
given by 
\begin{equation} \label{azimuthal}
    \left(\frac{d\phi}{d\tau}\right)^2 = \frac{\mathcal{L}^2}{g_{\phi\phi}^2} \, .
\end{equation}

Combining Eqs.~\eqref{radial} and \eqref{azimuthal} yields the trajectory equation:
\begin{eqnarray} \label{equatorial-motion3}
    \left(\frac{dr}{d\phi}\right)^2 
    = \frac{r^4}{\mathcal{L}^2} \left( \frac{B(r)}{A(r)} \mathcal{E}^2 
      - B(r)\left(1 + \frac{\mathcal{L}^2}{r^2}\right)\right) \, .
\end{eqnarray}

We now introduce the variable transformation
\begin{eqnarray}
    u = \frac{1}{r}, \qquad 
    \frac{du}{d\phi} = -\frac{1}{r^2}\frac{dr}{d\phi} \, .
\end{eqnarray} 
With this transformation, the trajectory equation \eqref{equatorial-motion3} takes the form
\begin{eqnarray}\label{eqGeodesic}
    \left(\frac{du}{d\phi}\right)^2 
    = \frac{B(u)\,\mathcal{E}^2}{A(u)\,\mathcal{L}^2} 
      - \frac{B(u)}{\mathcal{L}^2}\left(1+\mathcal{L}^2 u^2\right) \, ,
\end{eqnarray}
where the metric functions are rewritten as
\begin{align}
    A(u) &= (1-r_+ u)(1-r_- u)(1+r_* u)^2 \ , \nonumber \\
    B(u) &= \frac{(1-r_+ u)(1-r_- u)}{(1+r_* u)^2} \, .
    \label{Bu}
\end{align}

Differentiating Eq.~\eqref{eqGeodesic} with respect to $\phi$ provides the geodesic equation for the orbital motion of a massive test particle around the self-dual BH in LQG:
\begin{eqnarray}\label{geod-eq}
    \frac{d^2 u}{d\phi^2} &=& 
    \frac{r_+ \left(1+\mathcal{L}^2 u^2\right)(1-r_- u)(1+r_* u)^3 - 4 \mathcal{E}^2 r_*}{2 \mathcal{L}^2 (1+r_* u)^5} \nonumber\\
    &+& \frac{2 r_* \left(1+\mathcal{L}^2 u^2\right)(1-r_- u)(1-r_+ u)(1+r_* u)^2}{2 \mathcal{L}^2 (1+r_* u)^5} \nonumber\\
    &+& \frac{r_- \left(1+\mathcal{L}^2 u^2\right)(1-r_+ u)(1+r_* u)^3}{2 \mathcal{L}^2 (1+r_* u)^5} \nonumber\\
    &-& \frac{2 \mathcal{L}^2 u (1-r_- u)(1-r_+ u)(1+r_* u)^3}{2 \mathcal{L}^2 (1+r_* u)^5} \, .
\end{eqnarray}

\subsection{The perihelion shift}
Since $P$ is a small parameter and the geodesic equation (Eq.~\eqref{geod-eq}) is complicated, we expand its right-hand side and keep only the linear order. As a result, Eq.~\eqref{geod-eq} yields
\begin{align} \label{orbital-equation2}
    \frac{d^2 u}{d\phi^2} \approx \frac{M}{\mathcal{L}^2} - u + \frac{g(u)}{\mathcal{L}^2} \, ,
\end{align}
where
\begin{equation*}
    \frac{g(u)}{\mathcal{L}^2} = 3 M u^2 - \frac{4 M \big(\mathcal{E}^2 + 2 M u + 4 \mathcal{L}^2 M u^3\big) P}{\mathcal{L}^2} \, .
\end{equation*}

Following the approach of Ref.~\cite{Adkins_2007}, the perihelion shift after one complete revolution can be written as
\begin{equation}
    \Delta \varphi = \frac{\pi}{\mathcal{L}^2} \left| \frac{d g(u)}{du} \right|_{u = \frac{1}{R}}\, ,
\end{equation}
with $R = a(1-e^2)$, where $a$ and $e$ denote the semi-major axis and eccentricity of the orbit, respectively.  
In SI units, the quantities are redefined as
\begin{eqnarray}
    M \;\Rightarrow\; \frac{GM}{c^2}, 
\qquad 
L^2 \;\Rightarrow\; \frac{GM\,a(1-e^2)}{c^2} \, .
\end{eqnarray}
With these substitutions, the perihelion shift becomes
\begin{eqnarray} \label{eq:perishift}
    \Delta \varphi = 6\pi\alpha - 8\pi\alpha P - 48\pi\alpha^2 P \, ,
\end{eqnarray}
where
\begin{equation}
    \alpha = \frac{GM}{a c^2 (1-e^2)} \, .
\end{equation}

As an application to astronomical observations, we study the perihelion shift of Mercury, which serves as one of the most precise Solar System tests of gravitational theories.
The orbital parameters of Mercury are:
\begin{eqnarray*}
    \frac{2 G M_\odot}{c^2} &=& 2.95325008 \times 10^3 \, [\text{m}] \, , \nonumber \\
    a &=&  5.7909175 \times 10^{10} \, [\text{m}] \, , \nonumber \\
    e &=&  0.20563069\, .
\end{eqnarray*}

Substituting these values into Eq.~\eqref{eq:perishift}, the perihelion shift of Mercury is obtained as
\begin{equation} \label{numShift}
    \Delta \varphi = 2\pi\times(7.98744\times 10^{-8}-1.06499\times10^{-7} P ) \, .
\end{equation}

In Eq.~\eqref{numShift}, the first term represents the prediction of general relativity, whereas the second term arises from the correction due to the quantum correction parameter (polymeric function) $P$.  
For comparison, the measured perihelion advance of Mercury is given by \cite{Benczik02PRD,Iorio15IJMPD,Iorio2019ApJ,Shaymatov23ApJ}
\begin{equation}\label{obsShift}
    \Delta \varphi_{\text{obs}} = 2\pi \times (7.98734 \pm 0.00037) \times 10^{-8} \text{ rad/rev} \, .
\end{equation}
From Eqs.~\eqref{numShift} and \eqref{obsShift}, an upper bound on the quantum correction parameter $P$ for Mercury is obtained as
\begin{equation} \label{xi-mercury}
    P \leq 0.000043 \, .
\end{equation}
The influence of spacetime curvature on particle motion is most significant in the vicinity of strong gravitational sources, such as supermassive BHs. A well-known example is the system formed by the supermassive BH $\text{Sgr A}^*$ and its companion star S2 \cite{AbuterAmorim2020}. Since $\text{Sgr A}^*$ is about $10^5$ times more massive than S2, S2 star can be treated as a test particle moving in the strong gravitational field of the BH. According to high-precision astrometric measurements, the $\text{Sgr A}^*$--S2 system is characterized by the following parameters:
\begin{eqnarray*}
     \frac{2 G M_\odot}{c^2} &=& 2.95325008 \times 10^3 \, [\text{m}] \, ,\\
    M_{\text{Sgr A}^*} &=& 4.260 \times 10^6 M_{\odot}\, , \\
    a_{\text{S2}} &=& 970 \, [\text{au}] \, , \\
    1 \ \text{au} &=& 1.495978707\times10^{11} \, [\text{m}]\, , \\
    e_{\text{S2}} &=& 0.884649 \, , \\
    T_{\text{S2}} &=& 16.052 \, [\text{years}] \, .
\end{eqnarray*}
Using the parameters listed above, the perihelion shift of the S2 star is found to be
\begin{equation} \label{numShiftS2}
    \Delta \varphi = 48.298 -64.474\ P \Big[\ ^{\prime\prime}/\text{year} \Big] \, .
\end{equation}
For comparison, the observed perihelion shift of the S2 orbit around $\text{Sgr A}^*$ is reported as
\begin{equation} \label{obsShiftS2}
    \Delta \varphi = 48.298 \ f_{\text{SP}} \ \Big[\ ^{\prime\prime}/\text{year} \Big] \, ,
\end{equation}
with $f_{\text{SP}} = 1.10 \pm 0.19$ \cite{AbuterAmorim2020}.  
From Eqs.~\eqref{numShiftS2} and \eqref{obsShiftS2}, an upper bound on the quantum correction parameter $P$ for the S2 star is obtained as
\begin{equation} \label{xi-S2}
    P \leq 0.067419 \, .
\end{equation}
The upper limits obtained for $P$ are larger than those derived for Mercury in Ref.~\cite{TaoZhu2020PRD}, whereas the constraints determined for the S2 star are consistent with those reported in Ref.~\cite{TaoZhu2020PRD, TaoZhu2022JCAP}.
\begin{figure*}
\begin{tabular}{ccc}
  \includegraphics[scale=0.6]{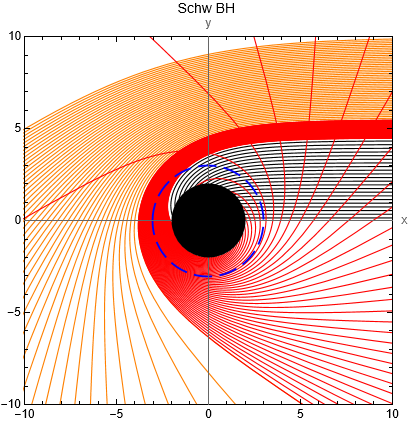}
  \includegraphics[scale=0.6]{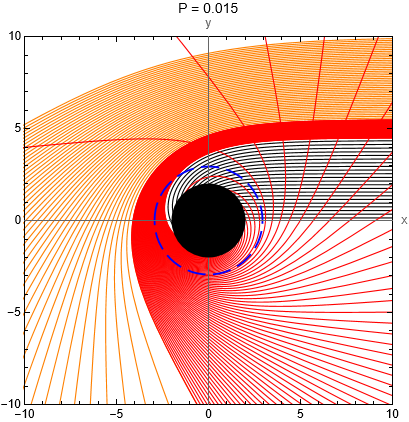}
  \includegraphics[scale=0.6]{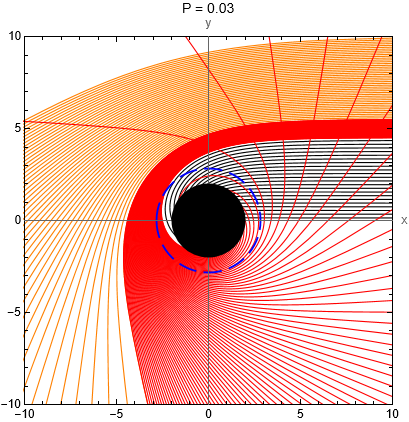}
  \end{tabular}
  \caption{The trajectories of photons are shown in separate plots for different values of $P$. Where the colors in all the plots are assigned as follows: black for $0 < b \leq 4.5$, red for $4.5 < b \leq 5.5$, and orange for $b > 5.5$. The blue dashed ring represents the photon sphere.}
\label{fig:ray1}
\end{figure*}
\begin{figure*}
   \centering
  \includegraphics[scale=0.61]{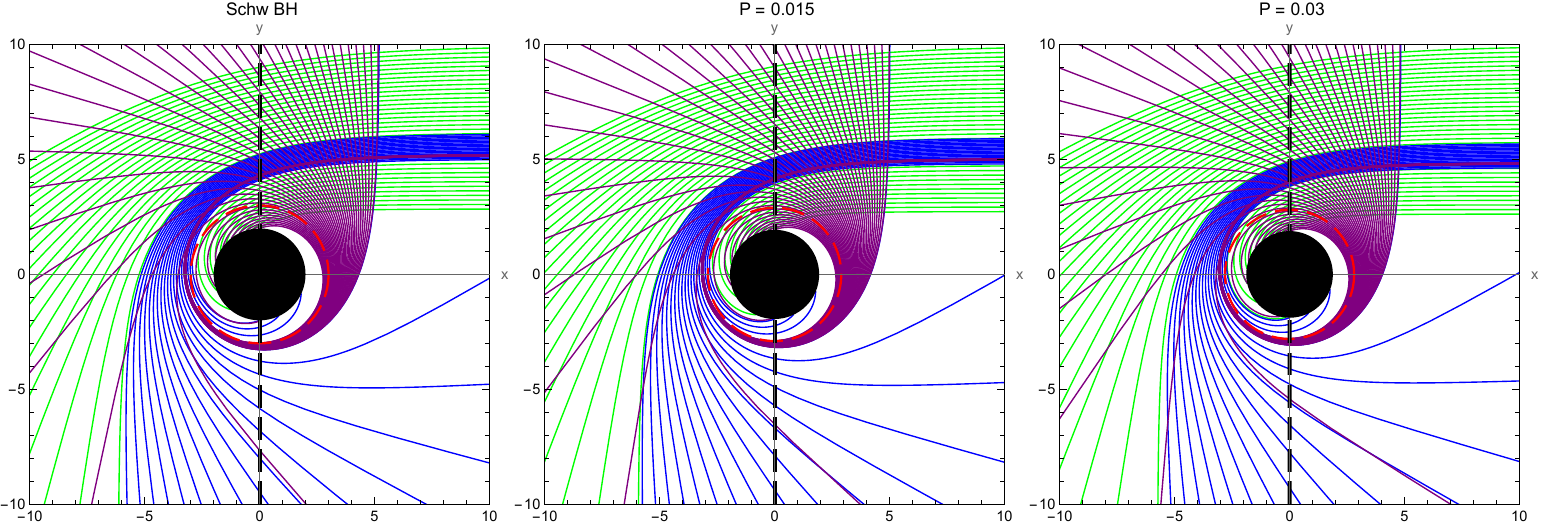}
 \caption{The trajectories of photons as a function of
$b$ for the self-dual BHs in LQG. The green, blue, and purple curves correspond to $1/4<n<3/4$, $3/4<n<5/4$, and $n>5/4$, respectively. The red dashed ring is the photon sphere.}
\label{fig:ray2}
\end{figure*}
\begin{figure} 
    \centering
    \includegraphics[scale=0.75]{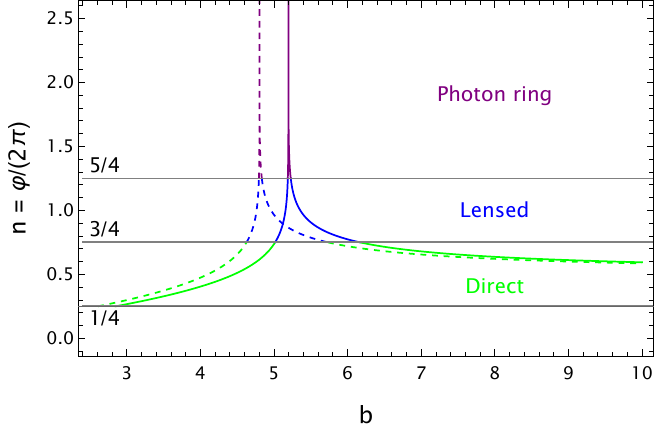}
    \caption{The plot shows the number of photon orbits $n$ as a function of the impact parameter $b$. The solid curve corresponds to the Schwarzschild BH, and the dashed curve represents the case with $P=0.03$.
    }
 \label{F_b1}
\end{figure}

\begin{table*}
\centering
\caption{Classification of photon trajectories by impact parameter $b$.}
\label{tab:nb}
\renewcommand{\arraystretch}{1.3}
\begin{tabular}{|c|c|c|c|c|}
\hline
\textbf{BHs} &
\textbf{Direct ($1/4 < n < 3/4$)} &
\textbf{Lensed ($3/4 < n < 5/4$)} &
\textbf{Photon ring ($n > 5/4$)} &
\textbf{$b_c$}\\
\hline
\textbf{SchwBH} &
$2.848 < b < 5.015$ \ \text{and} \ $b > 6.168$ &
$5.015 < b < 5.188$ \ \text{and} \ $5.228 < b < 6.168$ &
$5.188 < b < 5.228$\ & $5.196$ \\
\hline
\textbf{P=0.015} &
$2.725 < b < 4.816$ \ \text{and} \ $b > 5.944$ &
$4.816 < b < 4.986$ \ \text{and} \ $5.026 < b < 5.944$ &
$4.986 < b < 5.026$ & $4.994$ \\
\hline
\textbf{P=0.020} &
$2.686 < b < 4.752$ \ \text{and} \ $b > 5.873$ &
$4.752 < b < 4.921$ \ \text{and} \ $4.961 < b < 5.873$ &
$4.921 < b < 4.961$ & 4.929\\
\hline
\textbf{P=0.030} &
$2.610 < b < 4.627$ \ \text{and} \ $b > 5.733$ &
$4.627 < b < 4.794$ \ \text{and} $\ 4.835 < b < 5.733$ &
$4.794 < b < 4.835$ & 4.803\\
\hline
\end{tabular}
\end{table*}

\section{Null geodesics around a self-dual BH spacetime in LQG}\label{Null}

We then turn to consider the null geodesics around a self-dual BH spacetime in LQG. To this end, we first rewrite the four-momentum for a massless particle (photon) as follows:    
\begin{equation} \label{momentum-photon}
    p^\alpha = dx^\alpha/d\lambda \, .
\end{equation}  
As already mentioned, we confine the photon trajectory to the equatorial plane ($\theta=\pi/2$). With this in view, employing Eqs.~\eqref{ex-hamiltonian} and \eqref{momentum-photon} with its Hamiltonian satisfying $H=0$, we obtain the following equations of motion for light propagation:
\begin{equation}
    \frac{dt}{d\lambda}=\frac{E}{A(r)},
\end{equation}
\begin{equation}
    \frac{d\phi}{d\lambda}=\frac{L}{r^2},
\end{equation}    
\begin{equation}
    \frac{A(r)}{B(r)}\left(\frac{dr}{d\lambda}\right)^2=E^2-A(r)\frac{L^2}{r^2}.
\end{equation}
To simplify, we apply the transformation $\lambda'=L\lambda$, then rewrite the equations as
\begin{equation}
    \frac{dt}{d\lambda'}=\frac{1}{bA(r)},
\end{equation}
\begin{equation}
    \frac{d\phi}{d\lambda'}=\frac{1}{r^2}\ ,
\end{equation} 
\begin{equation}
    \frac{A(r)}{B(r)}\left(\frac{dr}{d\lambda'}\right)^2=\frac{1}{b^2}-\frac{A(r)}{r^2}=V_{eff}(r)\ ,
    \label{radialM}
\end{equation}
where $b=\frac{L}{E}$ is the impact parameter, and $V_{eff}$ is the effective potential.

The radial motion described by Eq. \eqref{radialM} can be expressed in terms of the azimuth angle $\phi$, as follows
\begin{equation}
    \frac{A(r)}{r^4 B(r)}\left(\frac{dr}{d\phi}\right)^2=\frac{1}{b^2}-\frac{A(r)}{r^2}=V_{eff}(r)\, .
    \label{radialPhi}
\end{equation}
The radius $r_{ph}$ of the photon sphere is determined by $\left. \frac{dV_{\text{eff}}}{dr} \right|_{r=r_{ph}}=0$. Solving the condition $V_{\text{eff}}(r_{ph})=0$ yields the critical impact parameter $b_c$:
\begin{equation}
b_c = \frac{r_{ph}}{\sqrt{A(r_{ph})}}.
\end{equation}
Transforming Eq. \eqref{radialPhi} via the substitution $r = 1/u$ yields the following equation
\begin{equation}
    \left( \frac{du}{d\phi} \right)^2 = \frac{B(u)}{A(u)b^2} - u^2 B\left(u \right)\equiv G(u).
\end{equation}
Using Eq.~\eqref{Bu}, we find the following equation
\begin{align}
    G(u)=&\frac{1}{b^2 u^4 \left(\frac{2 P}{(P+1)^2}+\frac{1}{u}\right)^4}\nonumber\\&-\frac{u^2 \left(\frac{1}{u}-\frac{2}{(P+1)^2}\right) \left(\frac{1}{u}-\frac{2 P^2}{(P+1)^2}\right)}{\left(\frac{2 P}{(P+1)^2}+\frac{1}{u}\right)^2}\ .
\end{align}

The total change in the azimuthal angle $\varphi$ is given by a piecewise function dependent on the impact parameter:
\begin{equation}
\varphi = 
\begin{cases} 
\displaystyle \ \ \
\int\limits_0^{u_h} \frac{du}{\sqrt{G(u)}}\ ,  \quad & b < b_c\  \\[12pt]
\displaystyle
2 \int\limits_0^{u_{\min}} \frac{du}{\sqrt{G(u)}}\ , \quad & b > b_c\ ,
\end{cases}
\end{equation}
where $u_h \equiv 1/r_h$ is the inverse of the outermost horizon radius $r_h$, $u_{min}$ is the smallest positive root of $G(u) = 0$.
The light rays arriving at a BH from an infinitely distant source are nearly parallel. Fig. \ref{fig:ray1} shows ray tracings near a self-dual BH in LQG, demonstrating how the rays are bent by different angles for different values of the impact parameter $b$. In all the plots, colors are designated as: black for $0 < b \leq 4.5$, red for $4.5 < b \leq 5.5$, and orange for values of $b$ greater than 5.5. For $P = 0.0$, 0.015, and 0.03, the calculated critical impact parameters are $b_c = 5.196$, 4.994, and 4.803, respectively. The associated figures show that photons with these parameters remain within the photosphere, as indicated by the blue dashed rings. As clearly shown in the figure, the decreasing deflection angle of the red lines indicates a weakening gravitational field as the quantum correction parameter $P$ increases.

 The photon trajectories produced near the BH can be divided into three classes: direct, lensed, and photon rings \cite{Gralla_2019}.
  The total number of orbits can be determined by the equation $n(b) = \varphi(b)/2\pi$. 
  To determine the boundaries of the three classes mentioned above, we use the following equation \cite{Peng_2021}
\begin{equation}
    n = \frac{2m-1}{4}, \quad m = 1, 2, 3, \cdots\ ,
\end{equation}
here, $m$ is the number of intersection points, that is, the number of times the rays cross the equatorial plane when the rays come in the direction of the polar axis.
In Fig. \ref{fig:ray2}, the number $m$ corresponds to the number of times a ray crosses the central vertical axis. For 
$1/4<n<3/4$, the ray crosses the axis once (direct). For 
$3/4<n<5/4$, it crosses twice (lensed). Finally, for $n>5/4$ , it crosses three or more times (photon ring).

For different BHs, the limiting values of the impact parameter $b$ for the direct, lensed, and photon rings are different. All these values are listed in Table~\ref{tab:nb} and correspond to the intersection points in Fig.~\ref{F_b1}. The peaks in this figure represent the critical impact parameter $b_c$ value. As the quantum correction parameter $P$ increases, the $b_c$ value decreases, indicating that the gravitational field diminishes.  Fig.~\ref{F_b1} corresponds to plots 1 and 3 in Fig.~\ref{fig:ray2}, and the same colors are used in both figures to represent the same ranges of the impact parameter $b$.

The trajectories can be categorized into three types \cite{Gralla_2019, Peng_2021}:
\begin{itemize}
    \item Direct trajectory: The rays are deflected by an angle $\varphi<3\pi/2$, with the number of orbits satisfying $1/4 < n < 3/4$. (the green rays in Fig.~\ref{fig:ray2}).
\item Lensed trajectory: the rays are deflected by an angle in the range $3\pi/2 < \varphi < 5\pi/2$ (the blue rays).
\item Photon ring trajectory: the number of orbits is $n > 5/4$ (the pink rays).
\end{itemize}

 Table~\ref{tab:nb} shows the classification of light rays moving around the self-dual BHs in LQG and their special case, the Schwarzschild BH, and their dependence on the impact parameter $b$. In the fourth column (photon ring), light crosses the vertical axis at least three times. The center of the interval for this column is defined by the critical impact parameter $b_c$. The number of orbits $n$ approaches infinity  ($n \to \infty$) as the impact parameter $b$ approaches this critical value  ($b \to b_c$). Furthermore, if $b < b_c$, the light is absorbed by the BH. In general, the larger the value of the quantum correction parameter $P$, the more the self-dual BH in LQG properties differ from those of a Schwarzschild BH. This means that the shadow ring and photon sphere of the self-dual BH in LQG become smaller. For example, the critical impact parameter and photon ring for a Schwarzschild BH are $b_c\approx 5.196$ and $5.188 < b < 5.228$, respectively. For a self-dual BH in LQG with $P=0.03$, these values are $b_c \approx 4.803$ and $4.794 < b < 4.835$.

%\cred{\section{}}

\section{IMAGES AND PHYSICAL PROPERTIES OF THIN ACCRETION DISK around self-dual BH in LQG}\label{sec:disk}

In this section, we study the radiation properties of an accretion disk composed of gas, dust, and plasma orbiting a self-dual BH in LQG. A widely used theoretical framework for describing such disks is the Novikov–Thorne model~\cite{Novikov:1973kta, Cui24}, which treats the disk as optically thick and geometrically thin. In this approximation, the vertical thickness $h$ is much smaller than the radial extent $r$, i.e., $h \ll r$, and can therefore be neglected in comparison to the disk’s radial size. From an observational standpoint, accretion disks are typically characterized by their electromagnetic flux, temperature distribution, and luminosity. In what follows, we provide a detailed analysis of the influence of the quantum correction parameter $P$ on the image, energy flux, and redshift.

\begin{figure*} 
    \centering
    \includegraphics[scale=0.55]{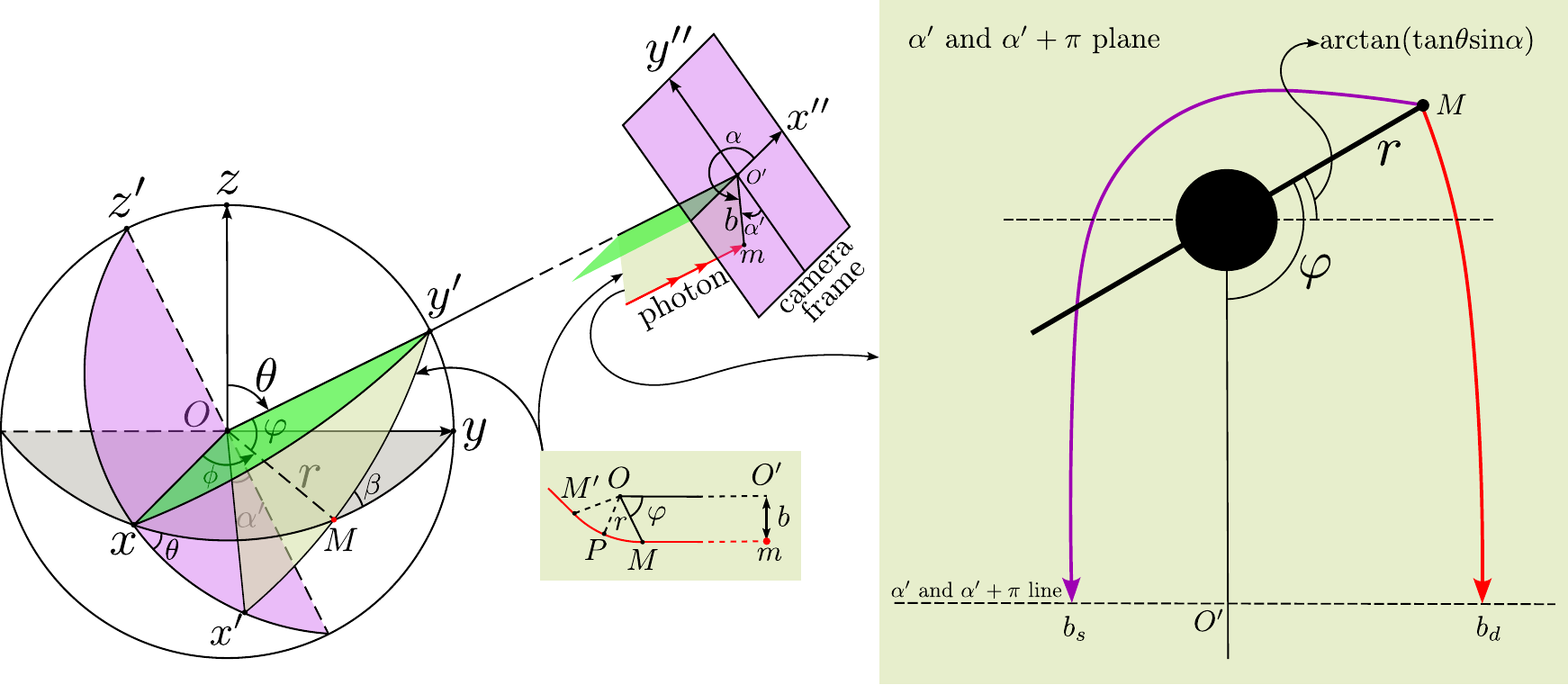}
    
    \caption{Schematic representation of the coordinate system used to construct the accretion disk image.
    }

 \label{fig:coordinate}
\end{figure*}

% \subsection{Observation coordinate system}
For the purpose of analyzing the image of a thin accretion disk, we make use of an observational coordinate system, as depicted in Fig.~\ref{fig:coordinate}. In this setup, the observer is placed at the position $(\infty, \theta, 0)$ in the spherical coordinate system of the BH $(r, \theta, \phi)$, where the origin $(r=0)$ is defined in the center of the BH.

Let us assume that, in the observer’s coordinate system $O'x''y''$ (the camera frame), a photon is emitted from the point $m(b,\alpha)$ and propagates perpendicularly to the camera plane toward the BH, where it intersects the disk at the point $M(r,\pi/2,\phi)$. According to the principle of optical path reversibility, a photon emitted from $M(r,\pi/2,\phi)$ on the accretion disk will follow a trajectory that arrives at the image point $m(b,\alpha)$ in the camera frame. In our analysis, we consider two types of images. The direct image $(b_d,\alpha)$ is generated by photons that propagate directly from the disk toward the observer’s camera. In contrast, the secondary image $(b_s,\alpha+\pi)$ arises from photons that initially move in the opposite direction but are bent around the BH by gravitational lensing before reaching the observer(see the right panel of Fig.~\ref{fig:coordinate}).

By applying the sine theorem of spherical triangles to $\triangle Myy'$ and $\triangle Mxx'$, we obtain the following relations:
\begin{eqnarray}
    \frac{\sin(\varphi)}{\sin(\frac{\pi}{2})} = \frac{\sin(\frac{\pi}{2}-\theta)}{\sin(\beta)}\, ,
\end{eqnarray}
\begin{eqnarray}
    \frac{\sin(\frac{\pi}{2}-\varphi)}{\sin(\theta)} = \frac{\sin(\frac{\pi}{2}-\alpha)}{\sin(\beta)}\, .
\end{eqnarray}
Using the above relations together with $\alpha+\alpha' = 3\pi/2$, the deflection angle $\varphi$ can be expressed as:
\begin{equation}
\varphi = \frac{\pi}{2} + \arctan(\tan\theta \sin\alpha).
\end{equation}

The angular deflections associated with the formation of the $n^{\text{th}}$-order image are denoted by $\varphi_n$ and are given by \cite{you2024}
\begin{widetext}
\begin{equation} \label{Eq:nthimage}
\varphi_n = 
\begin{cases}
\frac{n}{2} 2\pi + (-1)^n \left[ \frac{\pi}{2} + \arctan(\tan\theta \sin\alpha) \right], & \text{for $n$ even}, \\[6pt]
\frac{n+1}{2} 2\pi + (-1)^n \left[ \frac{\pi}{2} + \arctan(\tan\theta \sin\alpha) \right], & \text{for $n$ odd} \, ,
\end{cases} 
\end{equation}
\end{widetext}
where $n=0$ and $n=1$ correspond to the direct and secondary images, respectively.
\begin{figure*}[!htb]
    \centering
    \includegraphics[width=0.48\linewidth]{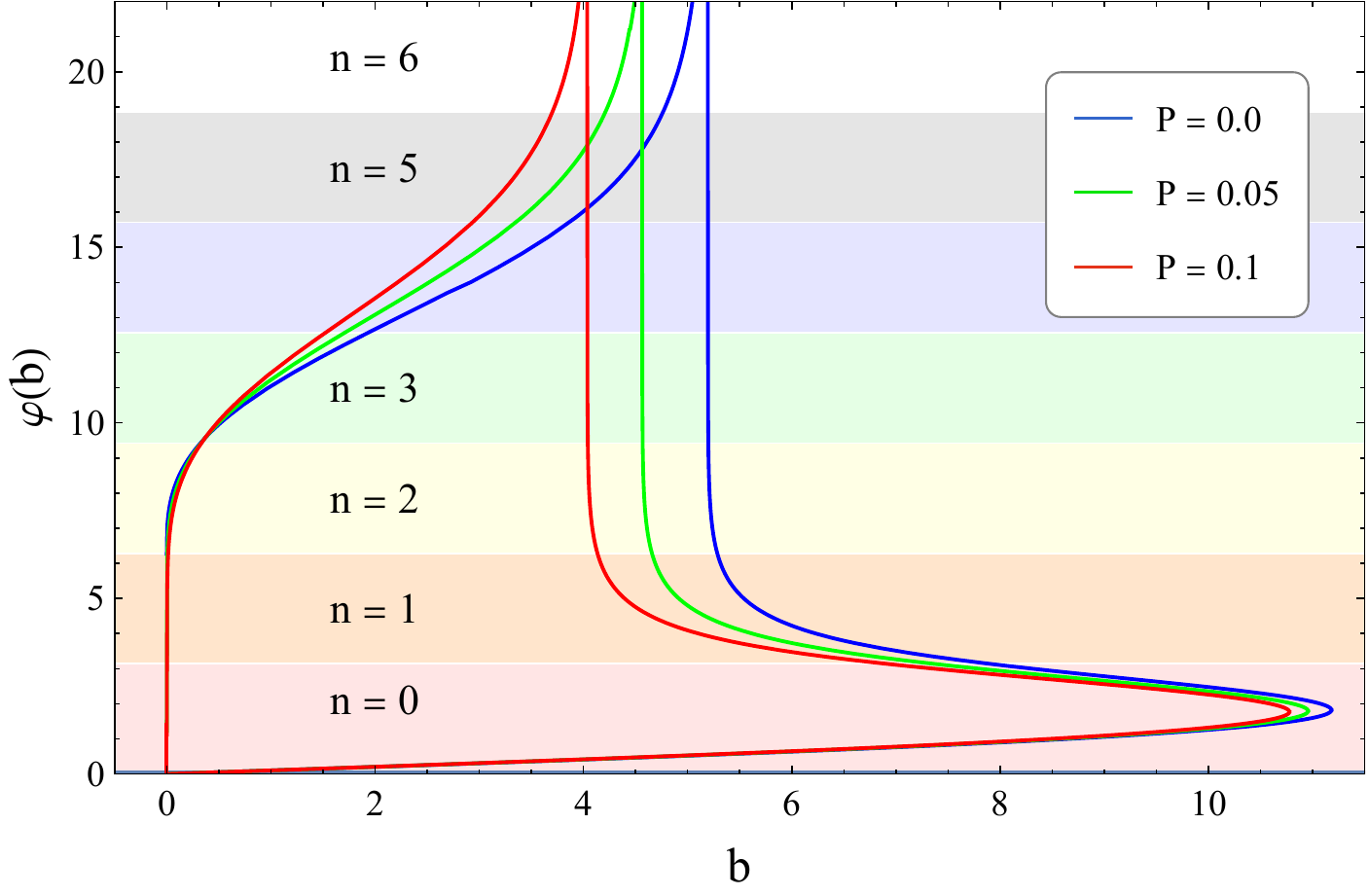}\hspace{0.5cm}
    \includegraphics[width=0.48\linewidth]{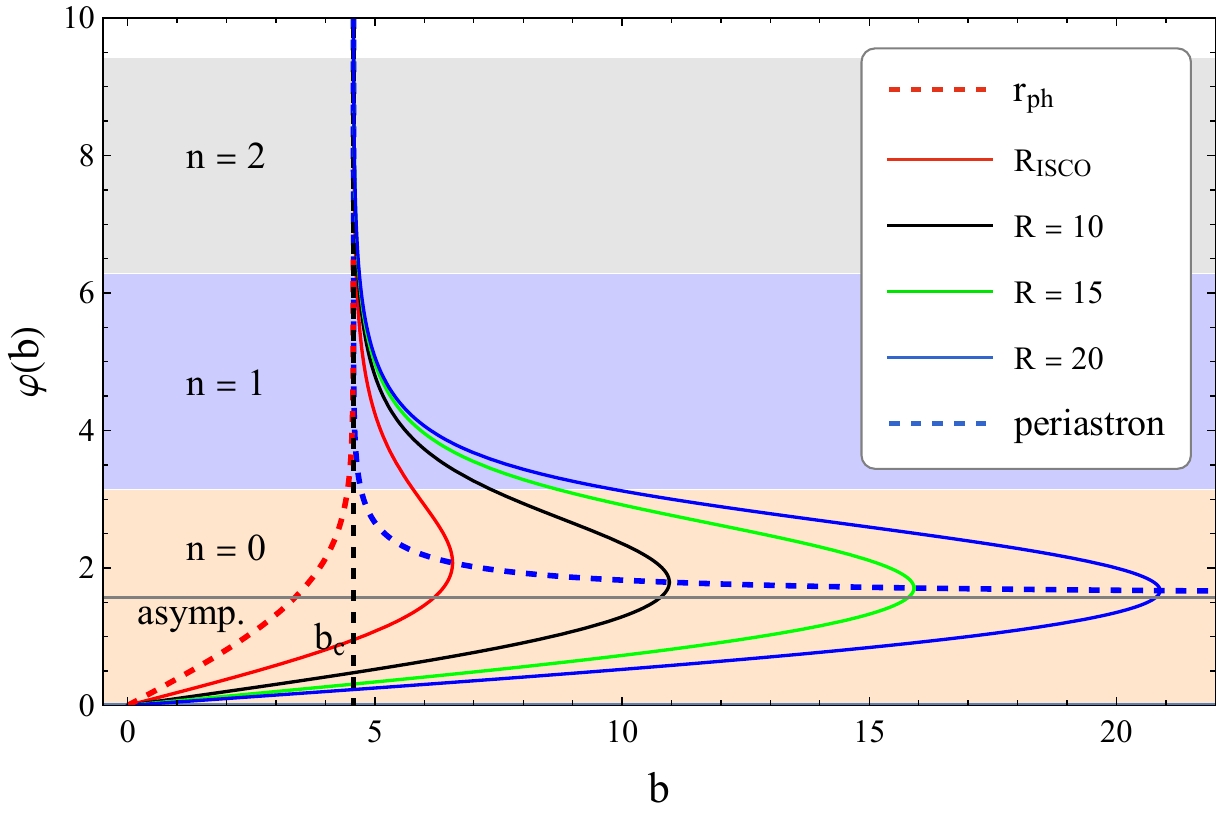}
    \caption{The image formation diagram for various values of the quantum correction parameter $P$ for the fixed radius $R=10$ (left panel) and of the radius $R$ for the fixed $P=0.05$ (right panel). Note that the innermost stable circular orbit refers to $R_{ISCO}=5.455$, corresponding to the solid red curve.}
    \label{fig:FBline1}
\end{figure*}
Suppose a photon is emitted from the camera plane at infinity with coordinates $(b,\alpha)$ and falls onto a circular orbit of radius $r$ in the BH accretion disk. In this case, the total deflection angle of the photon along its trajectory can be written as
\begin{equation} \label{Eq:phi1disk}
\varphi_1(b) = \int^{u_r}_{0} \frac{1}{\sqrt{G(u)}} du, 
\end{equation}
\begin{equation}\label{Eq:phi2disk}
\varphi_2(b) = 2 \int^{u_{\min}}_{0} \frac{1}{\sqrt{G(u)}} du - \int^{u_r}_{0} \frac{1}{\sqrt{G(u)}} du.
\end{equation}
Fig.~\ref{fig:FBline1} illustrates the dependence of the photon deflection angle $\varphi$ on the impact parameter $b$.
The left panel shows integral curves $\varphi_1$ and $\varphi_2$ for different values of the quantum correction parameter $P$. The peaks correspond to the critical values $b_c$ of the impact parameter. It is obvious that $b_c$ decreases significantly as $P$ increases. Interestingly, we observe that as $P$ increases, the impact parameter $b$ decreases significantly in $n\geq1$ region than in $n=0$, exhibiting that the decreasing rate of the secondary images is slightly larger than that of the direct images.
It should be mentioned that the curve for $P=0$ corresponds to a Schwarzschild BH case. From the right panel of Fig.~\ref{fig:FBline1}, the blue dashed curve corresponds to the deflection angle $\varphi_3$ of the photons at their periastron (i.e., the point of the closest approach to the BH for a given trajectory), which is determined by 
\begin{equation}
\varphi_3(b) = \int^{u_{\min}}_{0} \frac{1}{\sqrt{G(u)}} du\, . 
\end{equation}
Note that the blue dashed curve asymptotically approaches the line $\varphi=\pi/2$, below which the curves correspond to $\varphi_1$, while above it they correspond to $\varphi_2$. The intersection point of the curves defined by equations $\varphi_1$, $\varphi_2$, and $\varphi_3$ corresponds to maximum impact parameter $b_{max}$. As the radius $R$ increases, $b_{max}$ increases accordingly. The colored regions indicate the boundaries of the direct $(n=0)$ and $n^{\text{th}}$-order images. The accretion disk's radius starts at $R_{ISCO}$ and expands to the outer edge of the disk. For light to reach out from the larger radii of the disk $(R = 10, 15, 20)$, the rays must have the correspondingly larger impact parameters, as seen in the right panel of Fig.~\ref{fig:coordinate}. We find that the maximum impact parameters for these radii are $b_{max} =$ 10.96, 15.90 and 20.88, respectively. We can also observe that the secondary $(n \geq 1)$ images of the rays coming from various radii  remain almost unchanged compared to that of the direct image $(n = 0)$.

\begin{figure*}
\begin{tabular}{ccc}
  \includegraphics[scale=0.3]{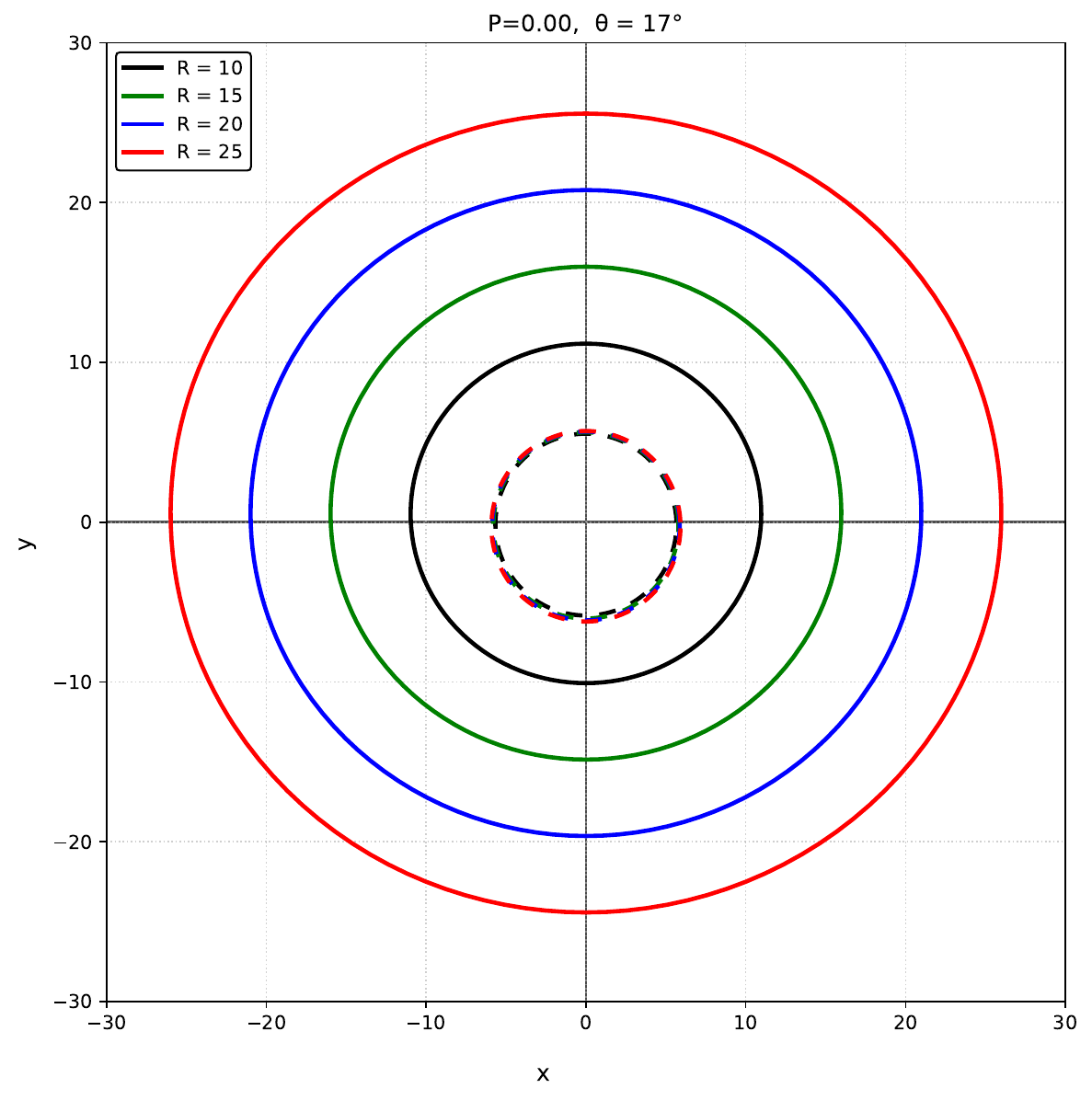}\hspace{-0.2cm}
  \includegraphics[scale=0.3]{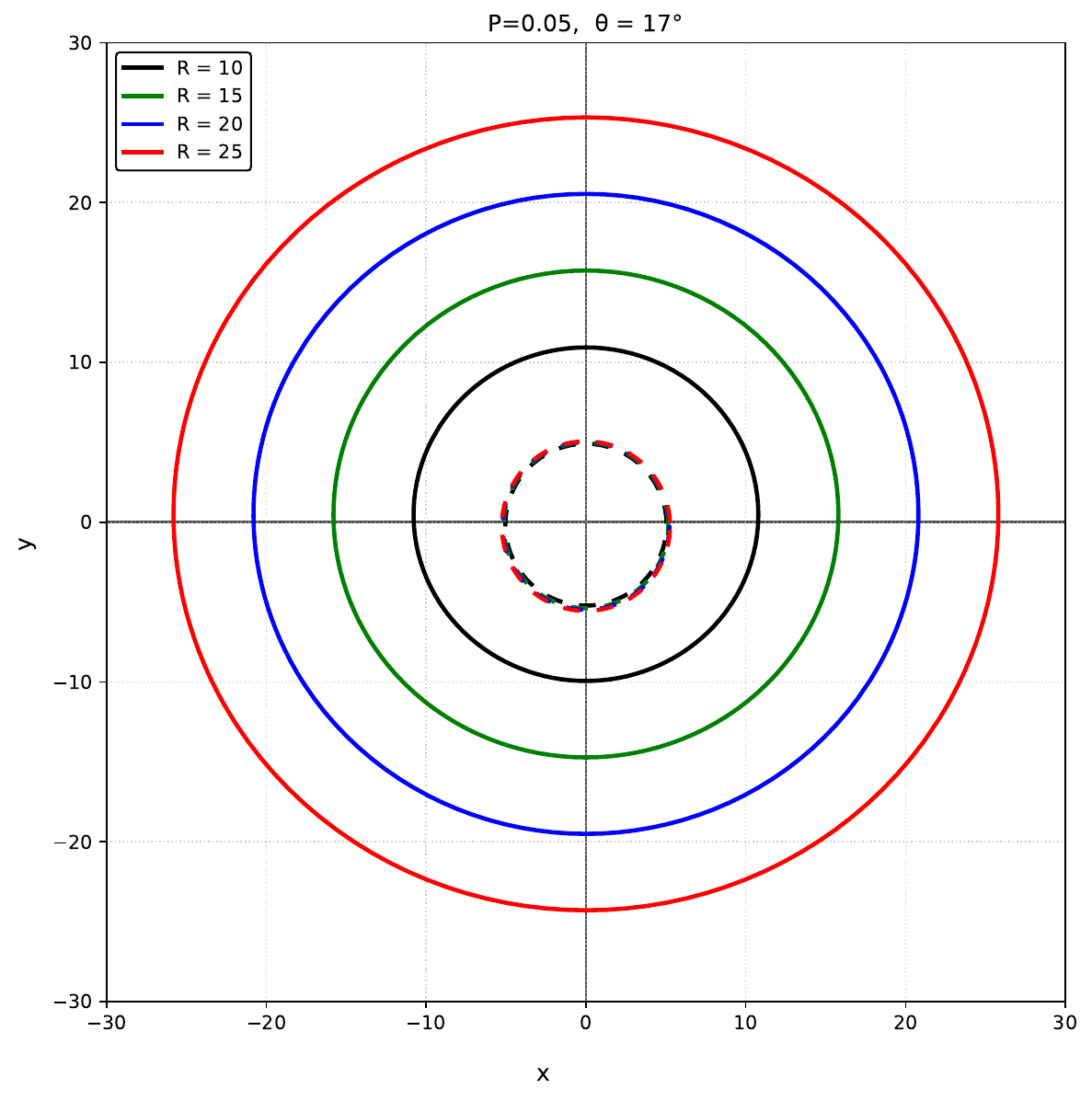}\hspace{-0.2cm}
  \includegraphics[scale=0.3]{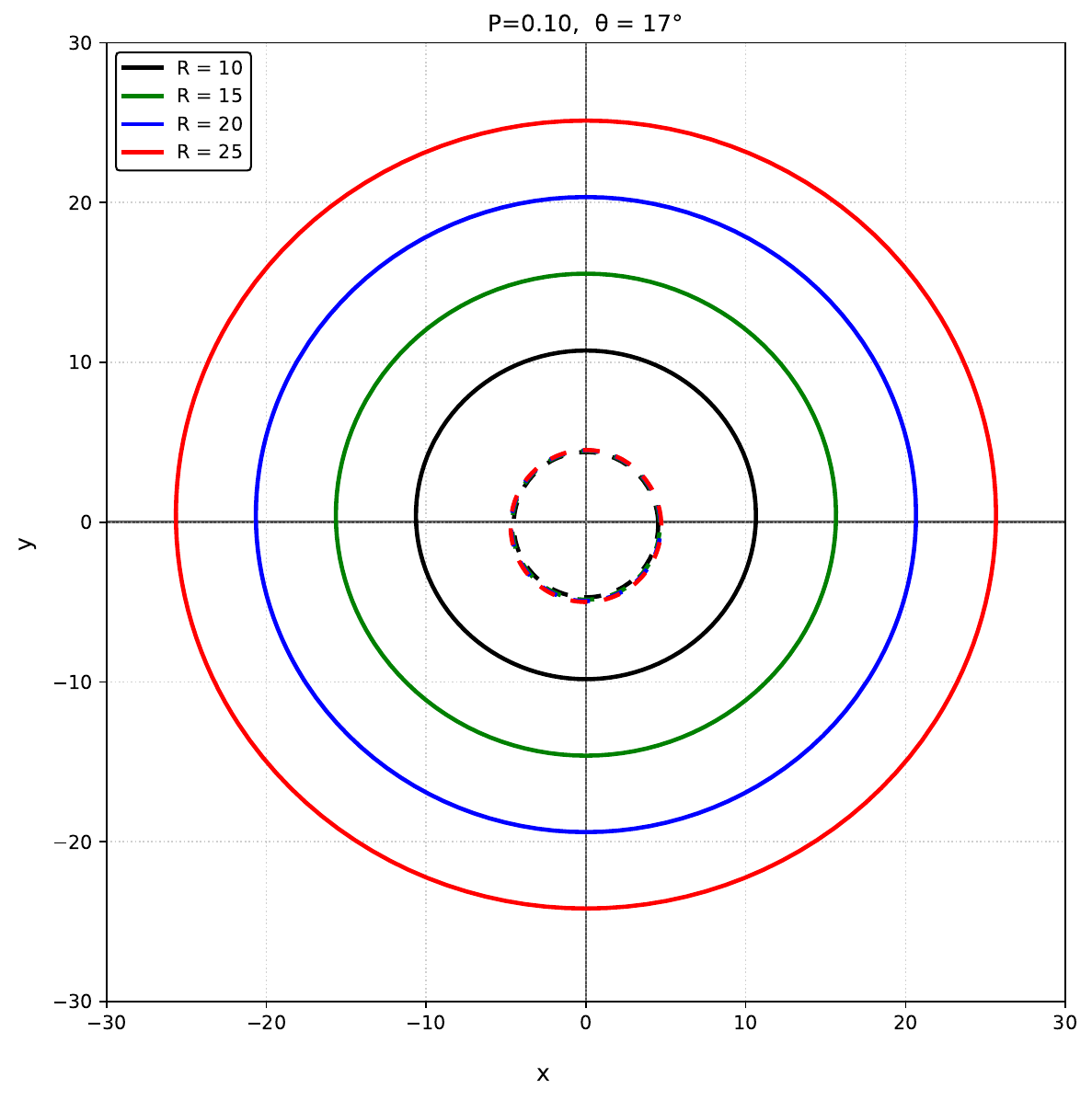}\\
  \includegraphics[scale=0.3]{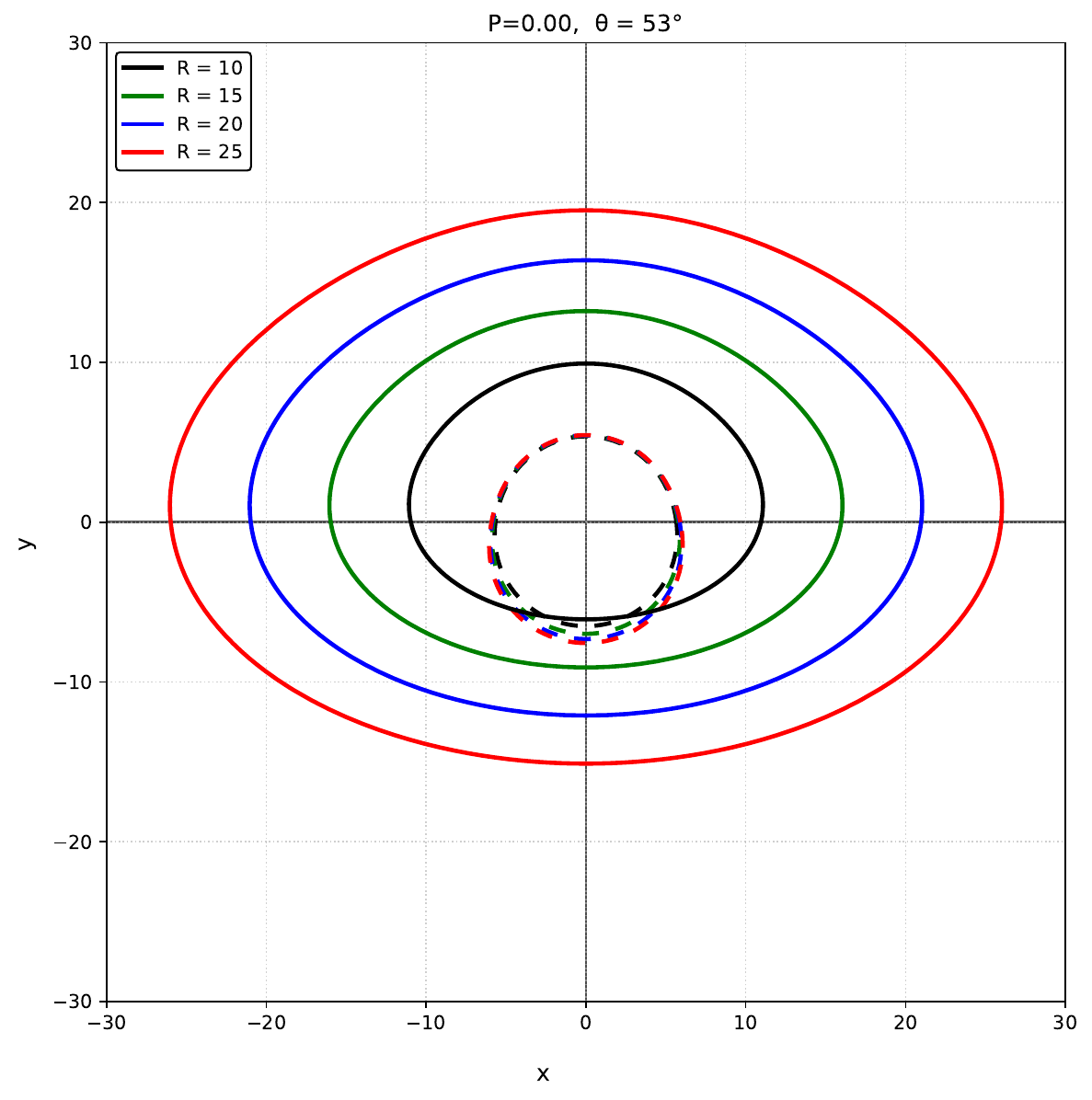}\hspace{-0.2cm}
  \includegraphics[scale=0.3]{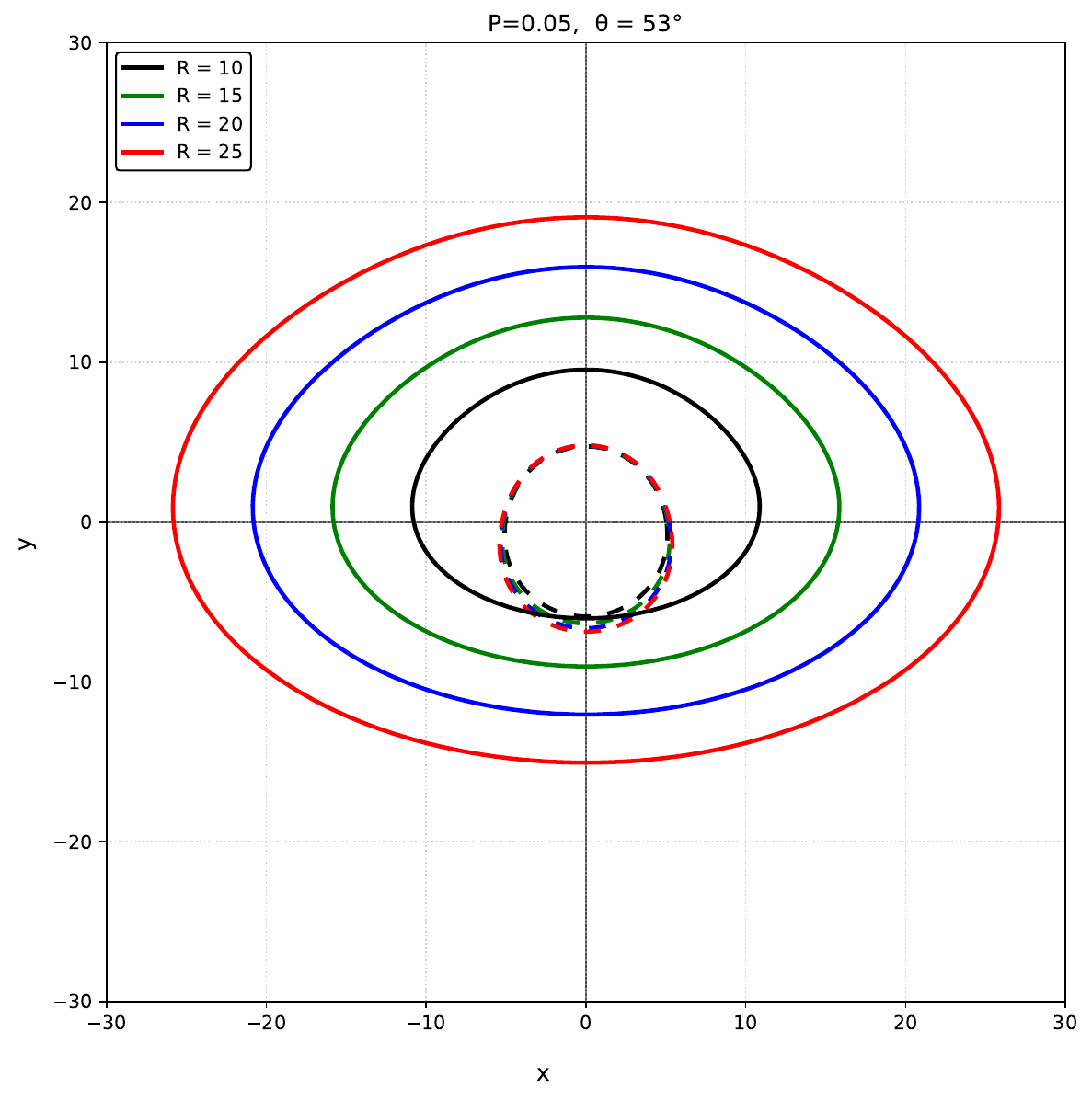}\hspace{-0.2cm}
  \includegraphics[scale=0.3]{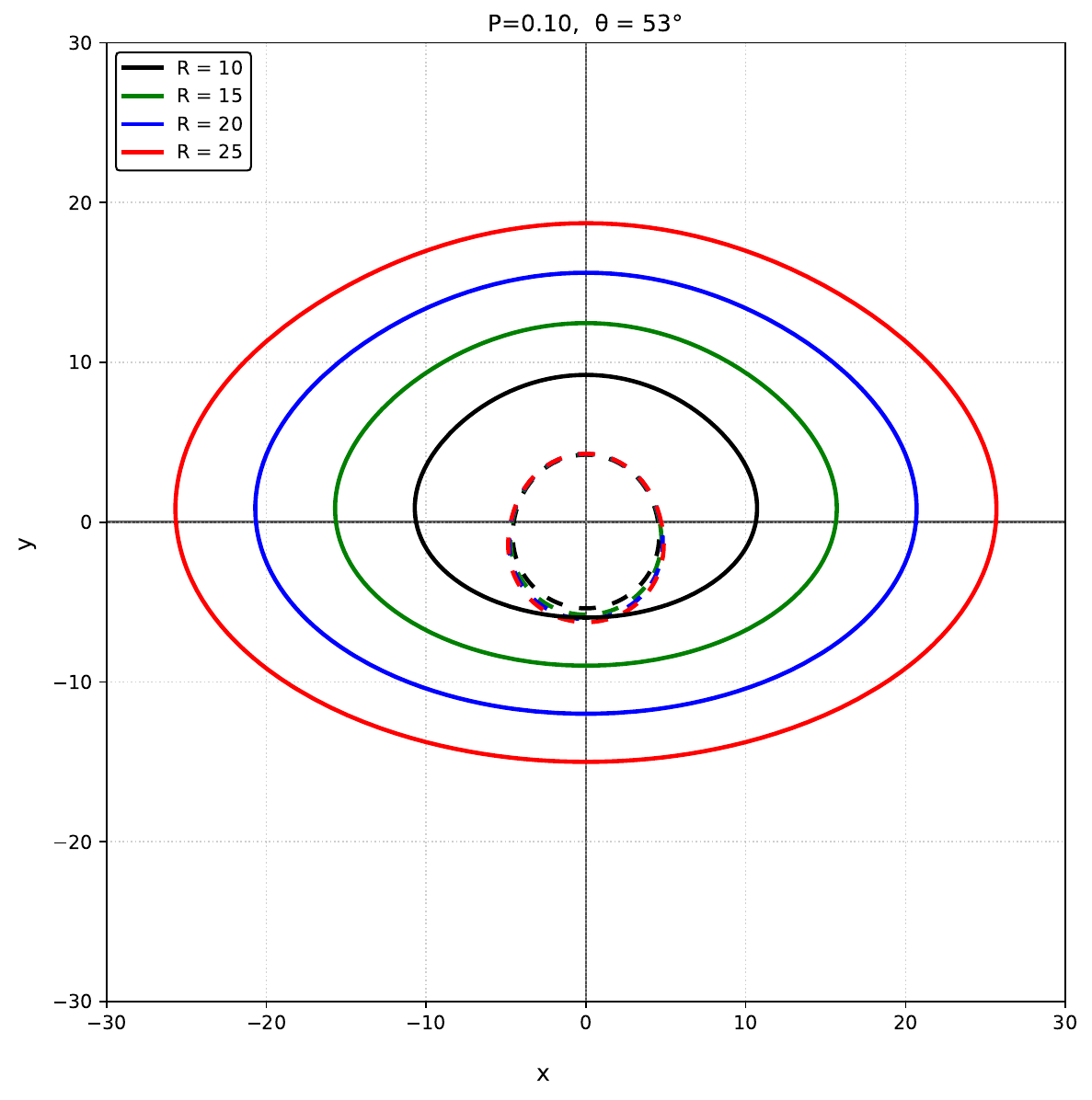}\\
  \includegraphics[scale=0.3]{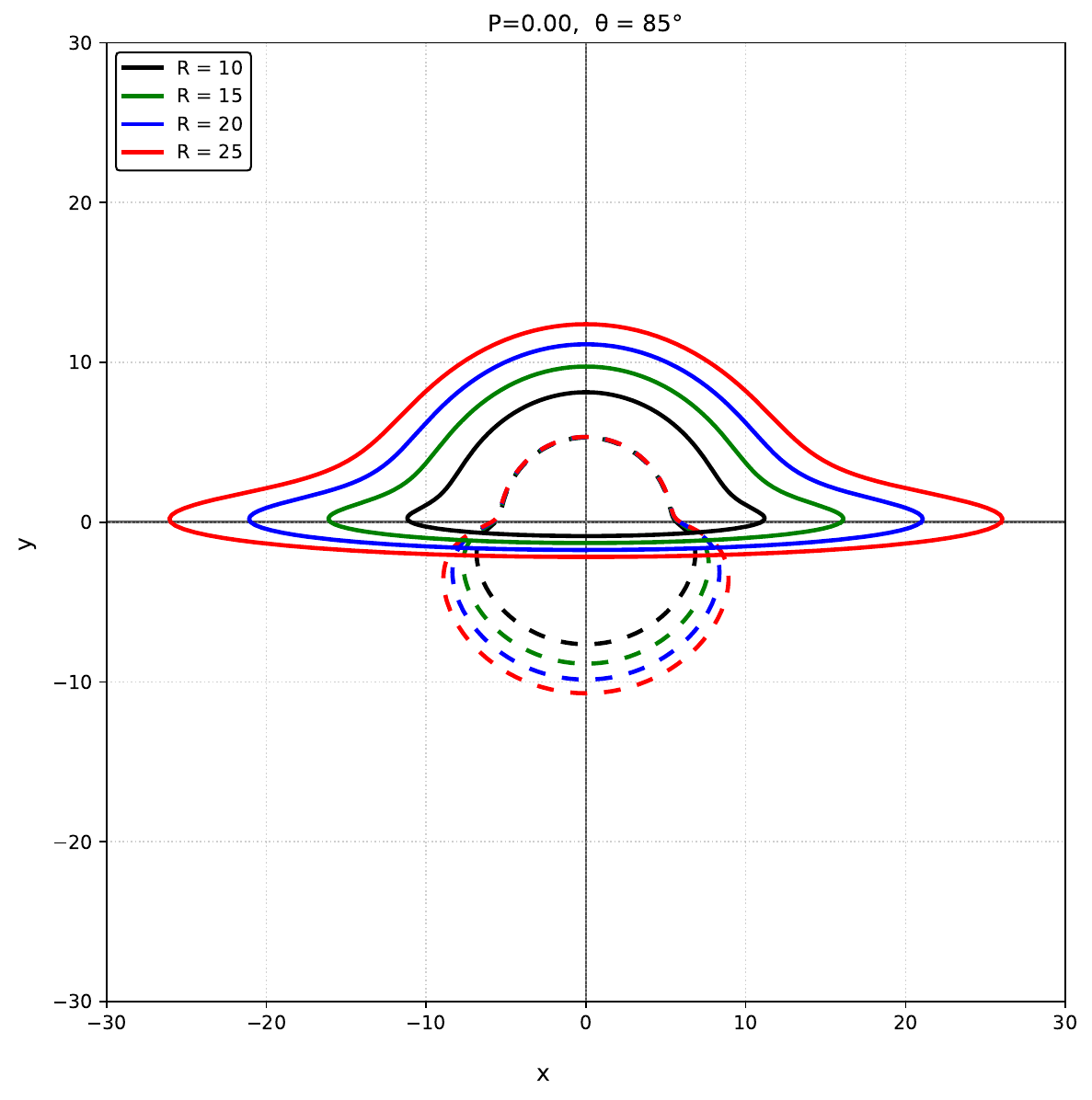}\hspace{-0.2cm}
  \includegraphics[scale=0.3]{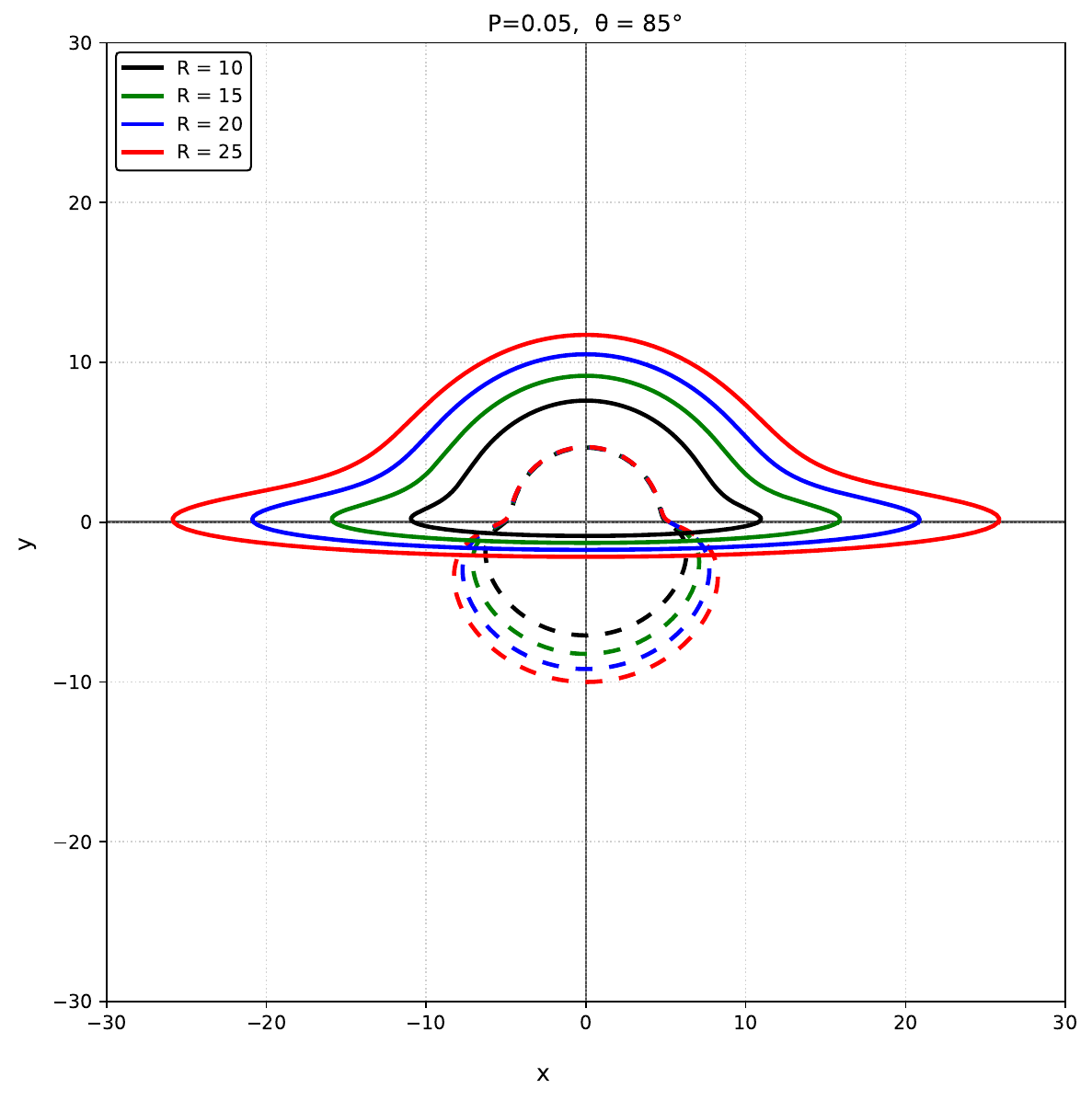}\hspace{-0.2cm}
  \includegraphics[scale=0.3]{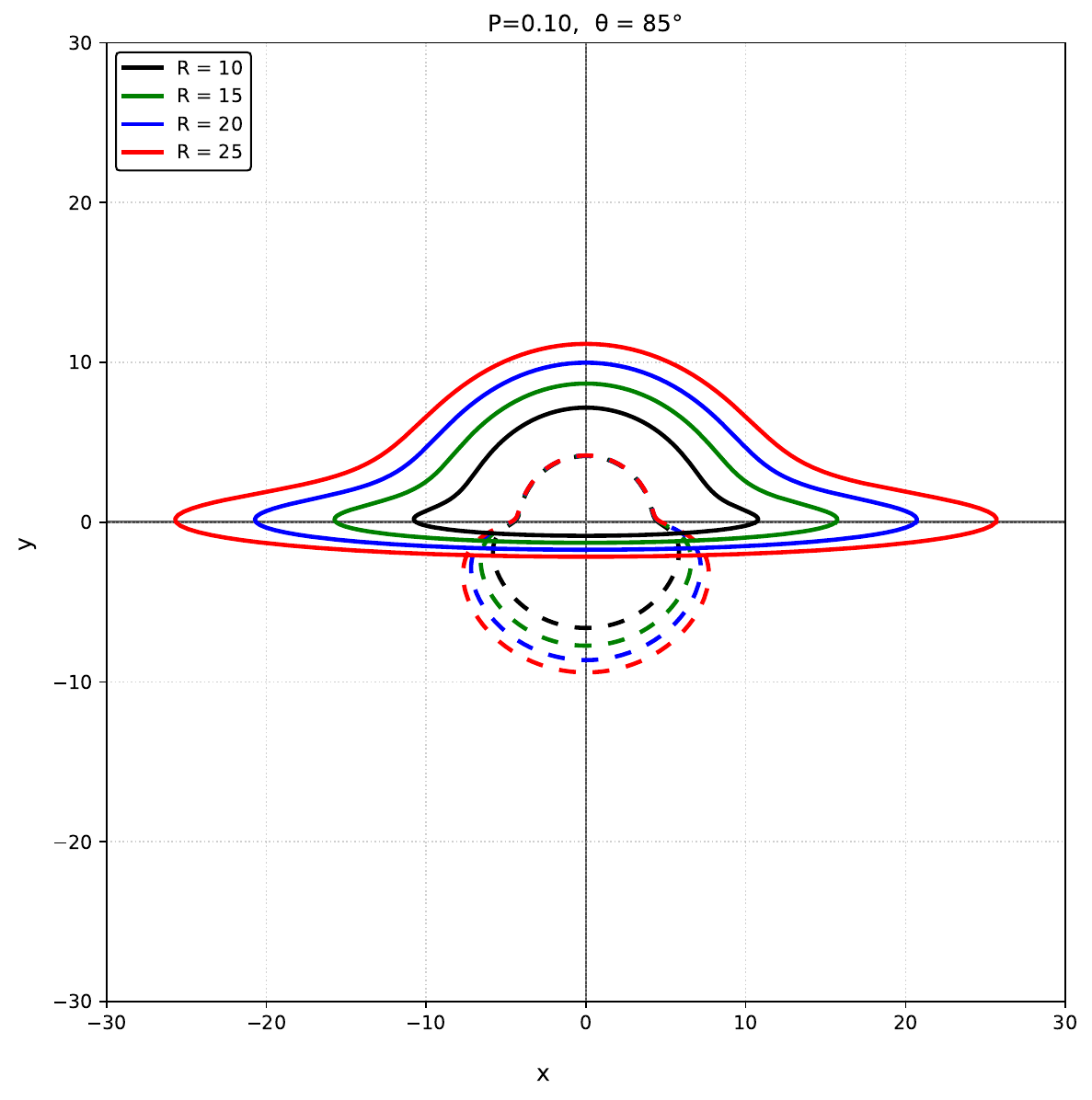}
  \end{tabular}
	\caption{\label{fig:accrthin} Direct and secondary image of the thin accretion disk around a self-dual BH in LQG. From top to bottom, the columns represent inclination angles of $17^\circ$, $53^\circ$, and $85^\circ$; from left to right, the rows correspond to $P$ values of $0.0$, $0.05$, and $0.10$. }
\end{figure*}

After numerically solving Eqs.~\eqref{Eq:nthimage}, \eqref{Eq:phi1disk}, and~\eqref{Eq:phi2disk}
 for the primary and secondary images, we obtain the coordinates $(b,\alpha)$ of the accretion disk in the observer's plane.
~\cite{Gyulchev20,Ziqiang25}. In Fig.~\ref{fig:accrthin}, the direct and secondary images of stable circular orbits around a self-dual BH in LQG, as observed by a distant observer, are depicted for different inclination angles. From Fig.~\ref{fig:accrthin}, it is evident that increasing the quantum correction parameter $P$ (left to right along each row) leads to a reduction in the size of both the primary and secondary images. This implies that with increasing $P$, a smaller impact parameter $b$ is required for photons to reach the same stationary circular orbit, indicating a weakening of the gravitational field.

The surface radiant energy flux is given by \cite{Novikov1973,Shakura:1972te,Thorne:1974ve}
\begin{equation}
F(r) = -\frac{\dot{M}_0 \Omega_{,r}}{4\pi \sqrt{-g}(E - \Omega L)^2}
\int_{r_{\text{isco}}}^{r} (E - \Omega L)L_{,r}dr\, , 
\end{equation}
where $\dot{M}_0$ represents the mass accretion rate, $g$ is the determinant of the spacetime metric, and $E$, $L$, and $\Omega$ correspond to the energy, angular momentum, and angular velocity of a particle on a circular orbit, respectively. For a particle in a circular orbit around a static, spherically symmetric BH, the quantities $E$, $L$, and $\Omega$ are defined as follows \cite{Shapiro83,Shaymatov22a}:
\begin{equation}
E = -\frac{g_{tt}}{\sqrt{-g_{tt} - g_{\phi\phi}\Omega^2}}, 
\end{equation}
\begin{equation}
L = \frac{g_{\phi\phi}\Omega}{\sqrt{-g_{tt} - g_{\phi\phi}\Omega^2}}, 
\end{equation}
\begin{equation}
\Omega = \frac{d\phi}{dt} = \sqrt{-\frac{g_{tt,r}}{g_{\phi\phi,r}}}.
\end{equation}

\begin{figure}
    \centering
    \includegraphics[width=1\linewidth]{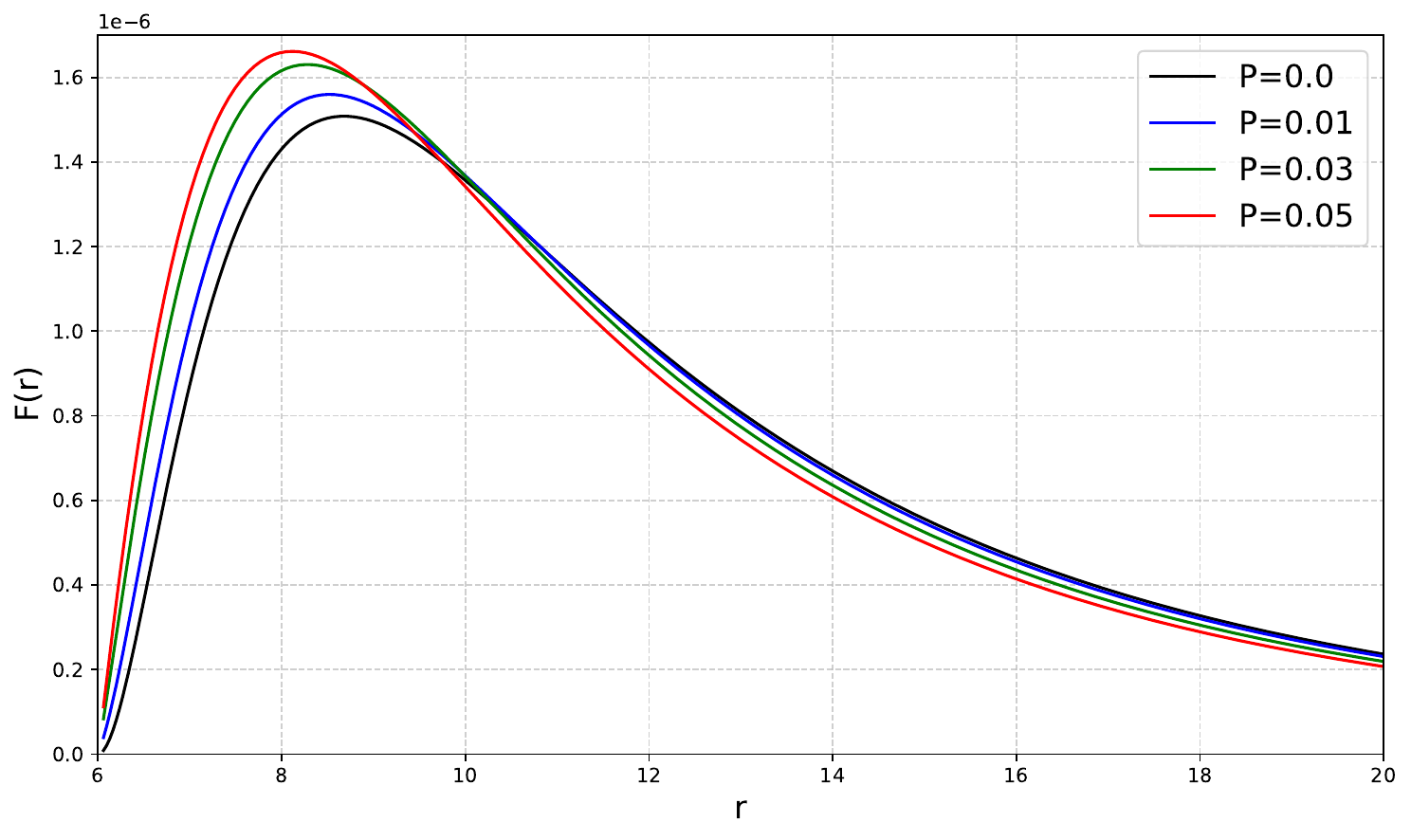}
    \caption{The energy flux $F(r)$ as a function of the radius $r$ for different values of  the quantum correction parameter $P$.}
    \label{fig:FluxFR}
\end{figure}

Fig.~\ref{fig:FluxFR} shows the energy flux $F(r)$ of a self-dual BH in LQG as a function of the radius $r$ for different values of the quantum correction parameter $P$. The figure shows that increasing the quantum correction parameter $P$ enhances the peak of the energy flux $F(r)$ and shifts it toward smaller radii.

At infinity, the observed radiation flux is influenced by both gravitational redshift and Doppler effects. As shown in Ref.~\cite{Ellis2009}, it is given by:  
\begin{equation}
F_{\text{obs}} = \frac{F(r)}{(1+z)^4}, 
\end{equation}
where $z$ is the redshift factor, expressed as~\cite{1979A&A}
\begin{equation}
1 + z =
\frac{1 + \Omega b \sin \theta \cos\alpha}{\sqrt{-g_{tt} - g_{\phi\phi}\Omega^2}}. 
\end{equation}

\begin{figure*}
\begin{tabular}{ccc}
  \includegraphics[scale=0.35]{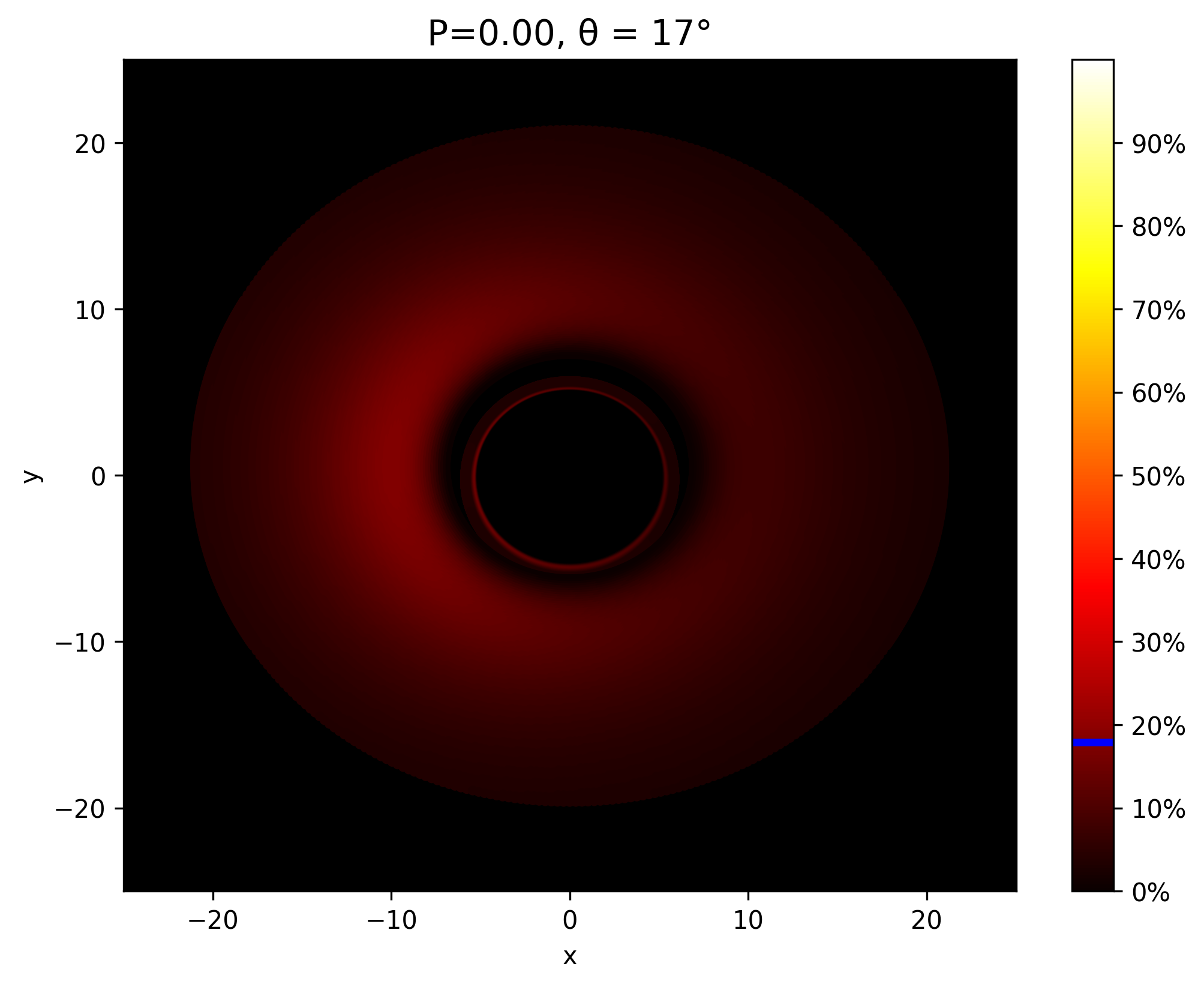}\hspace{-0.2cm}
  \includegraphics[scale=0.35]{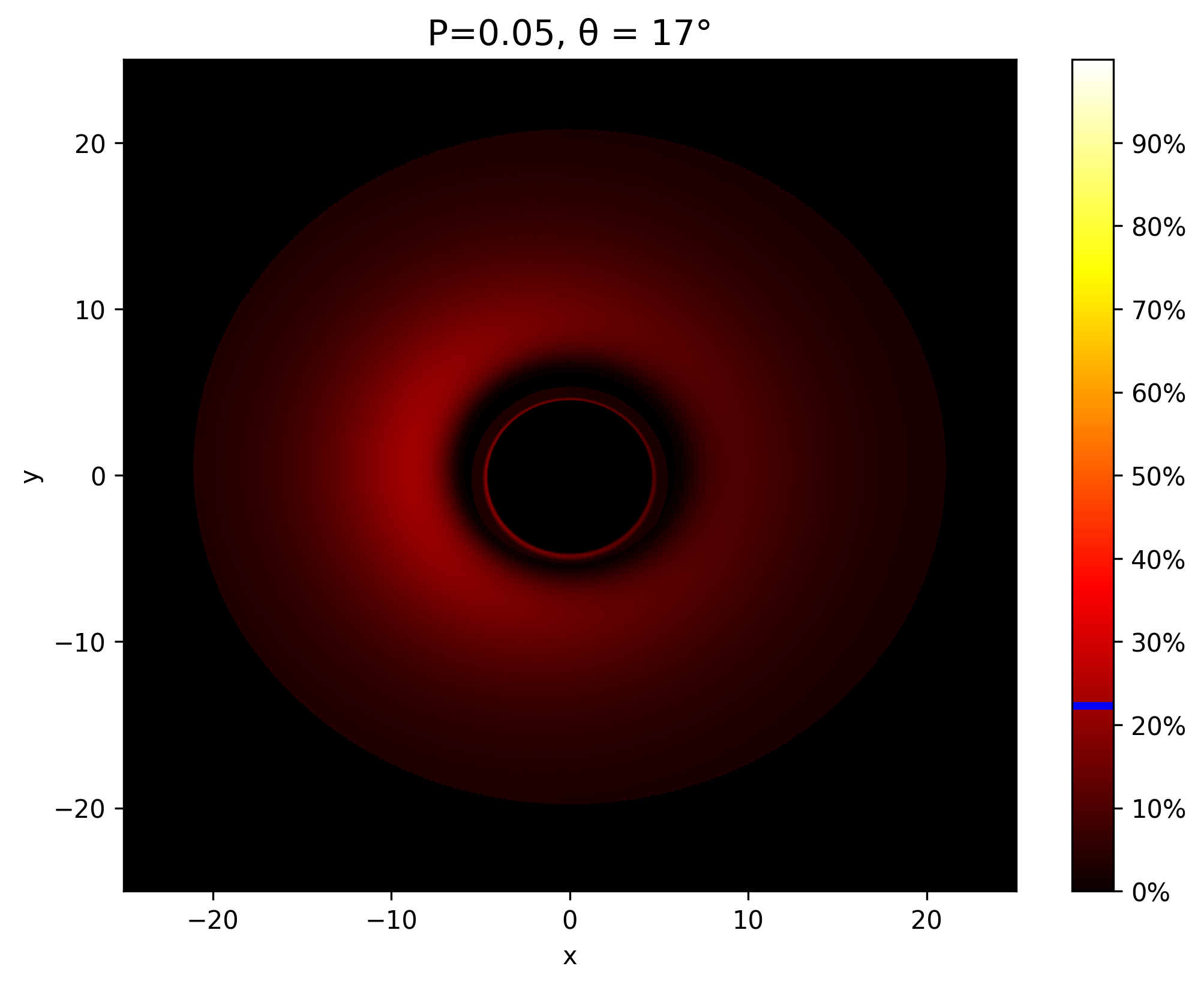}\hspace{-0.2cm}
  \includegraphics[scale=0.35]{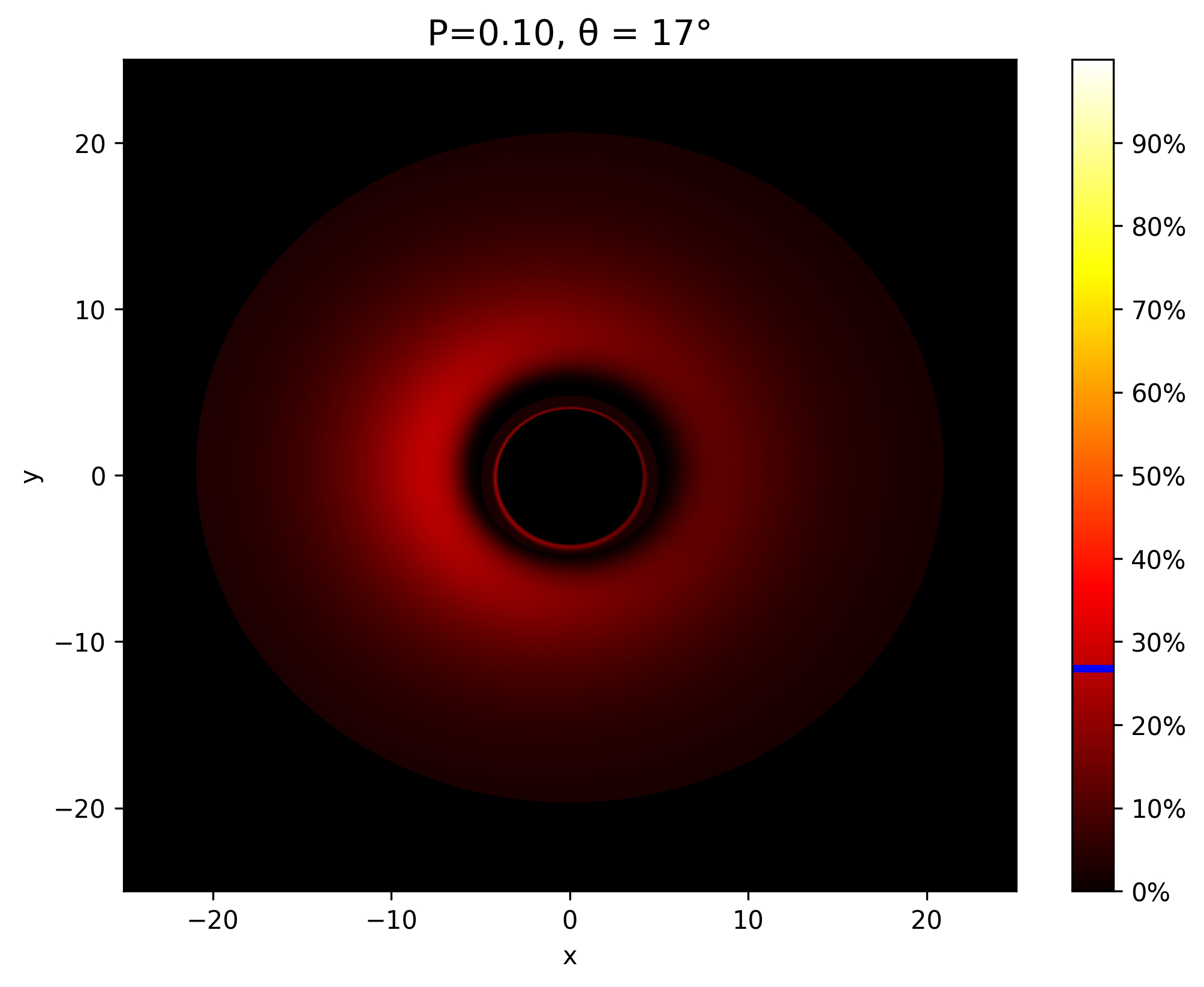}\\
  \includegraphics[scale=0.35]{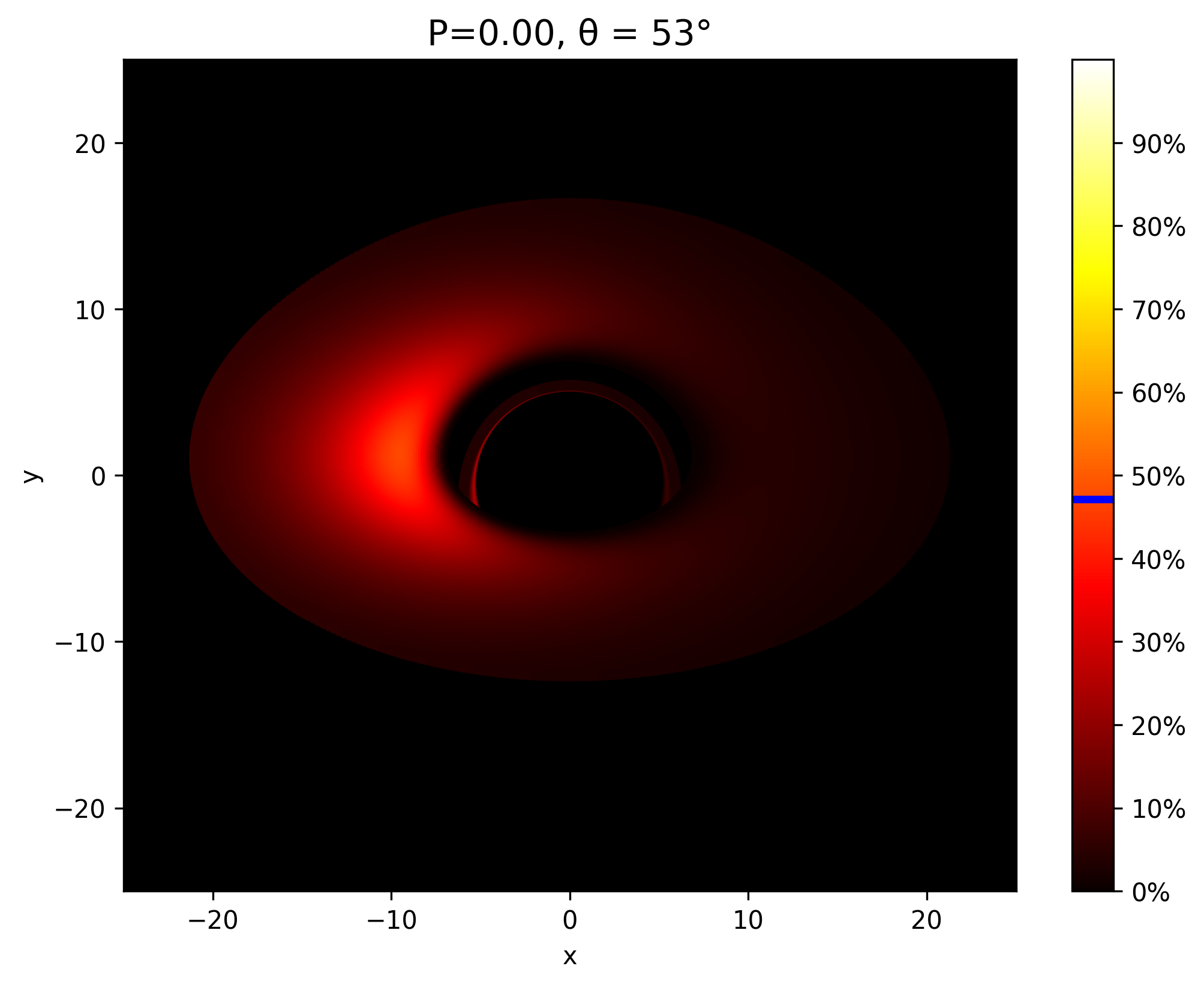}\hspace{-0.2cm}
  \includegraphics[scale=0.35]{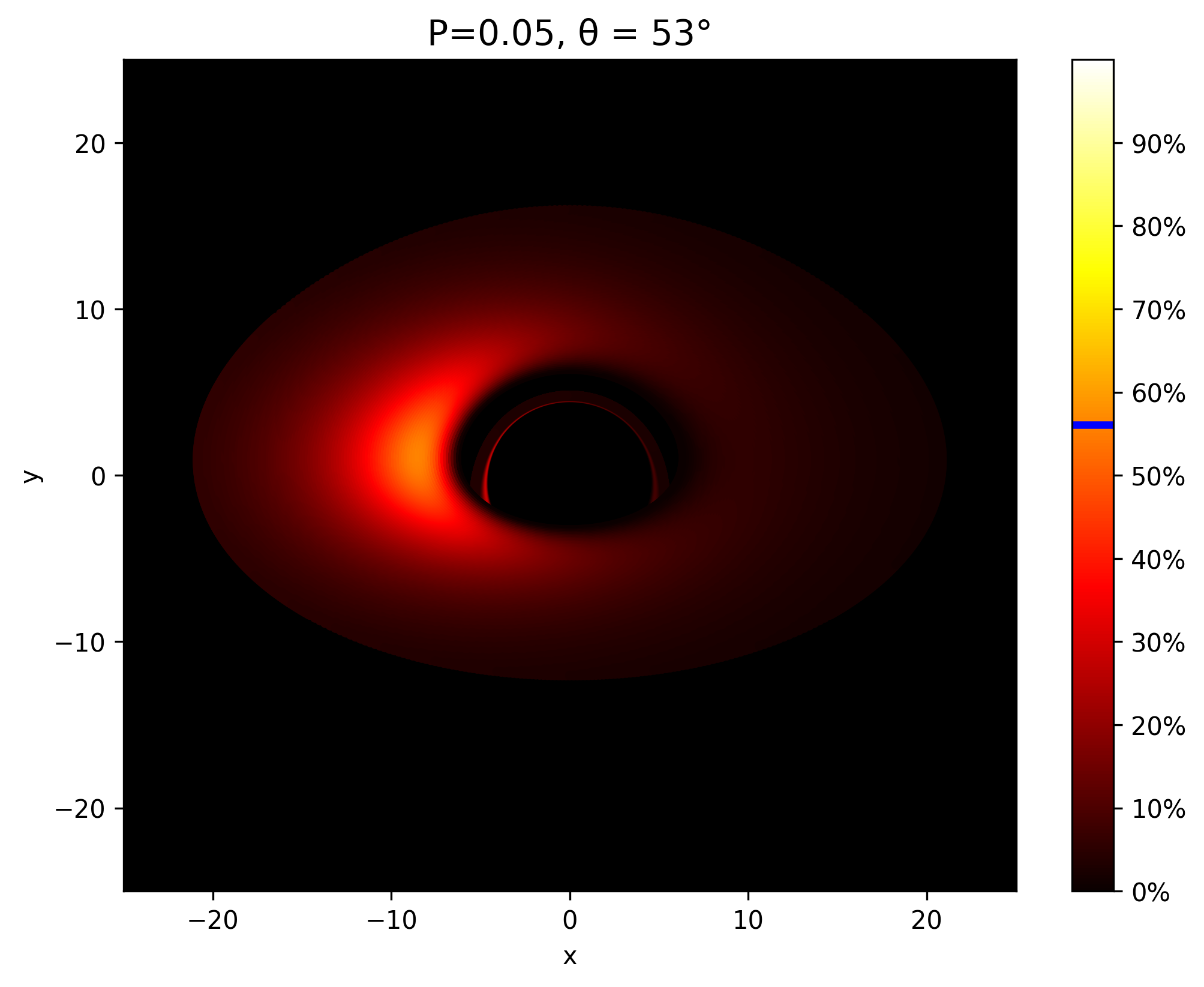}\hspace{-0.2cm}
  \includegraphics[scale=0.35]{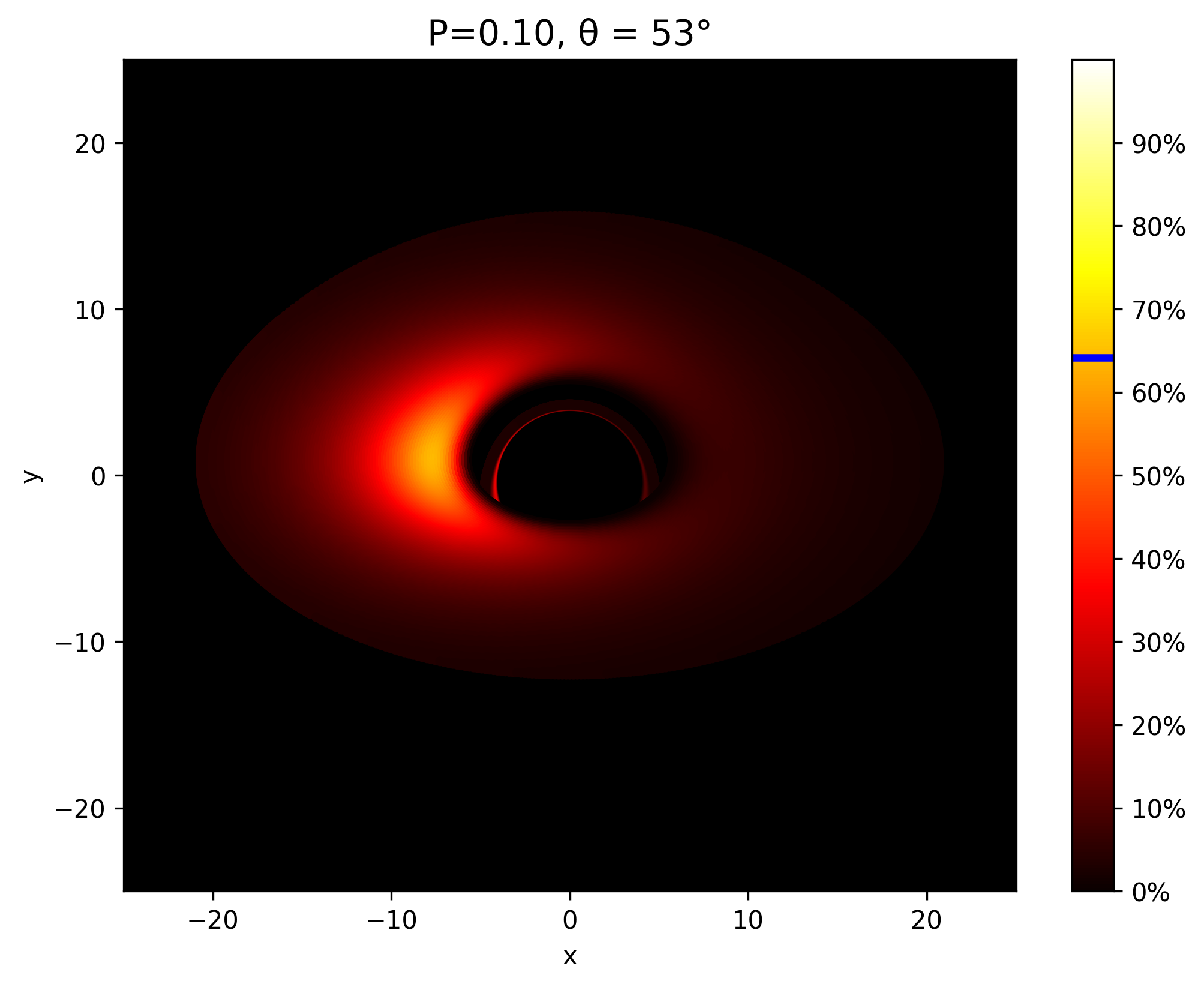}\\
  \includegraphics[scale=0.35]{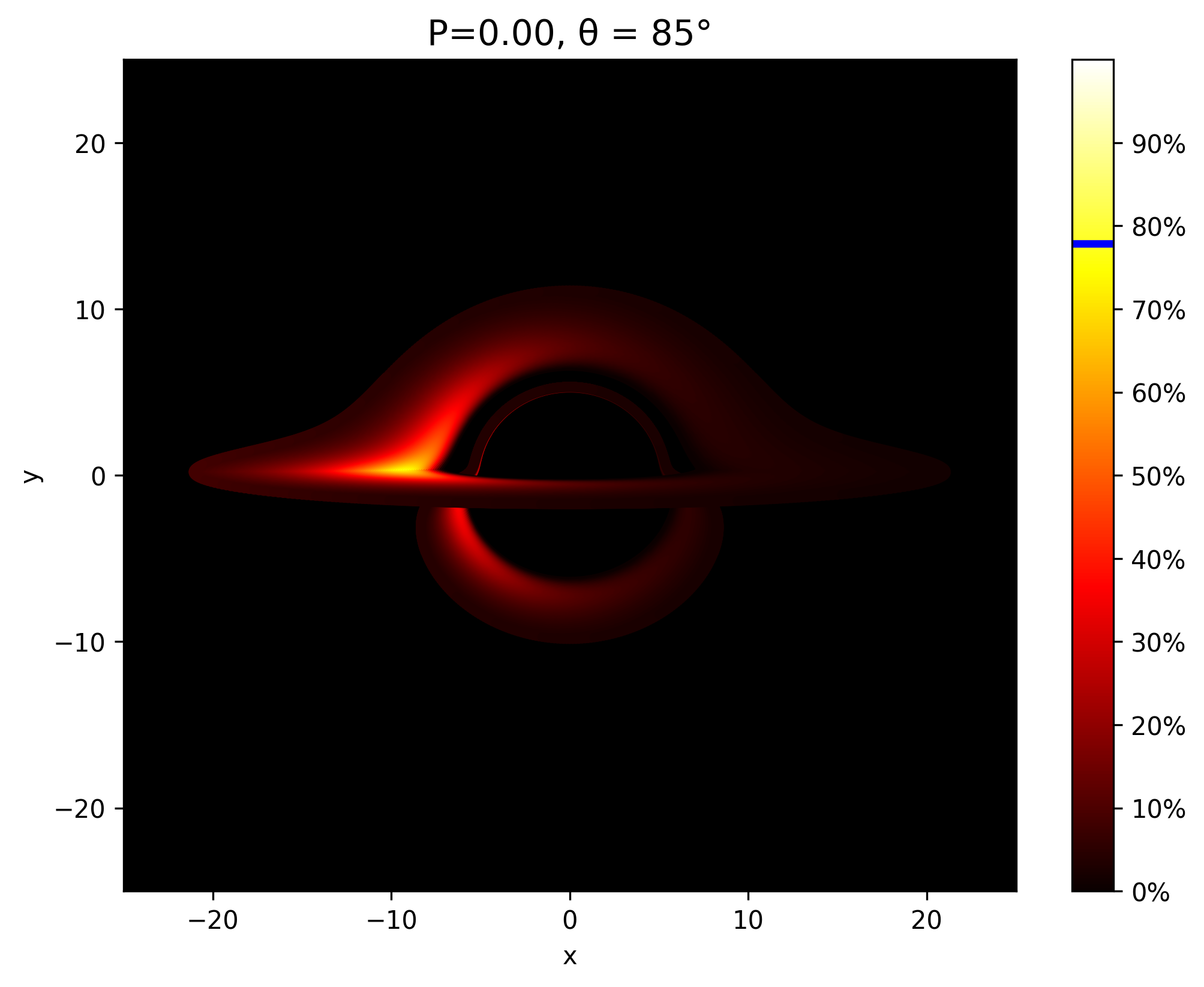}\hspace{-0.2cm}
  \includegraphics[scale=0.35]{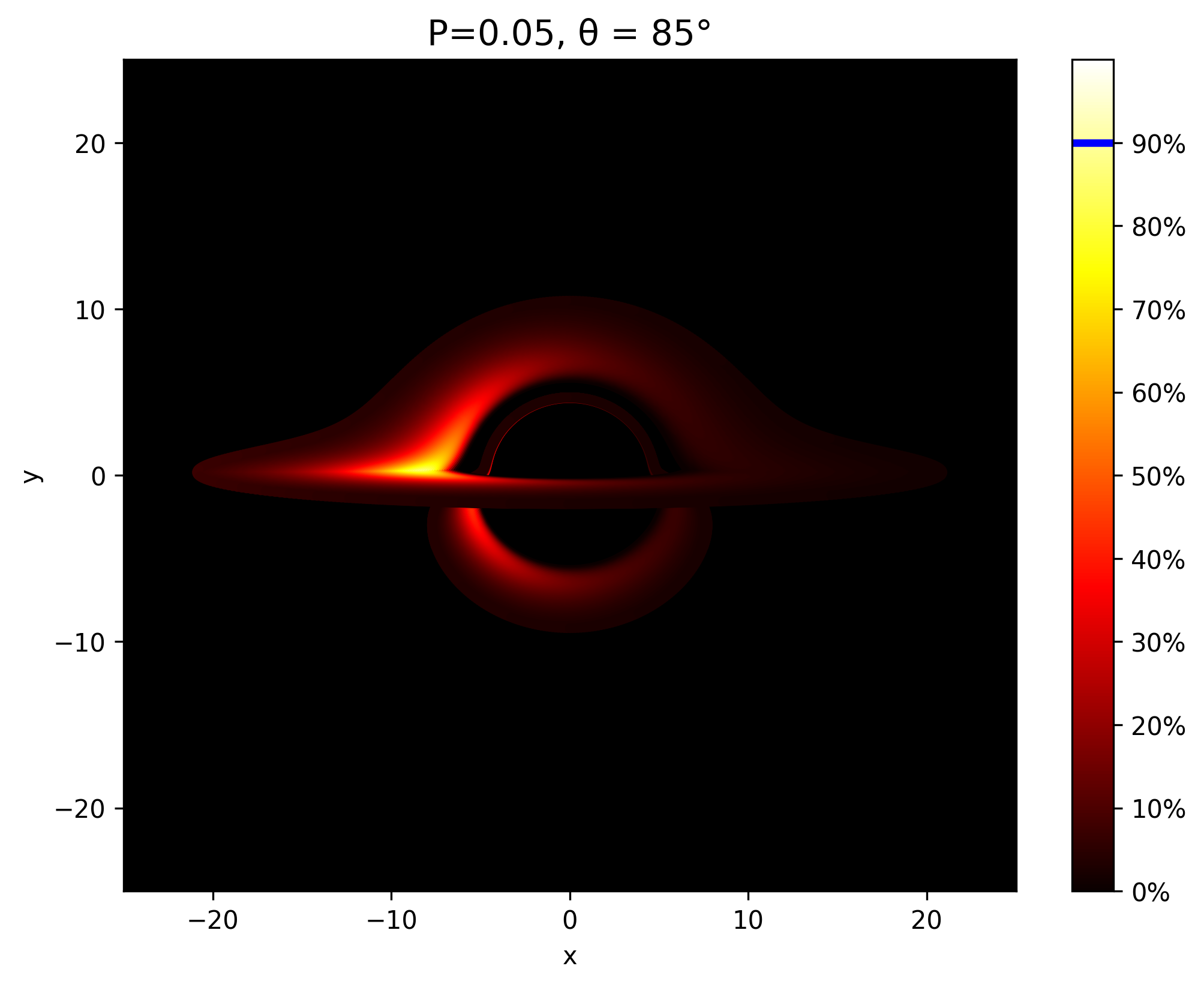}\hspace{-0.2cm}
  \includegraphics[scale=0.35]{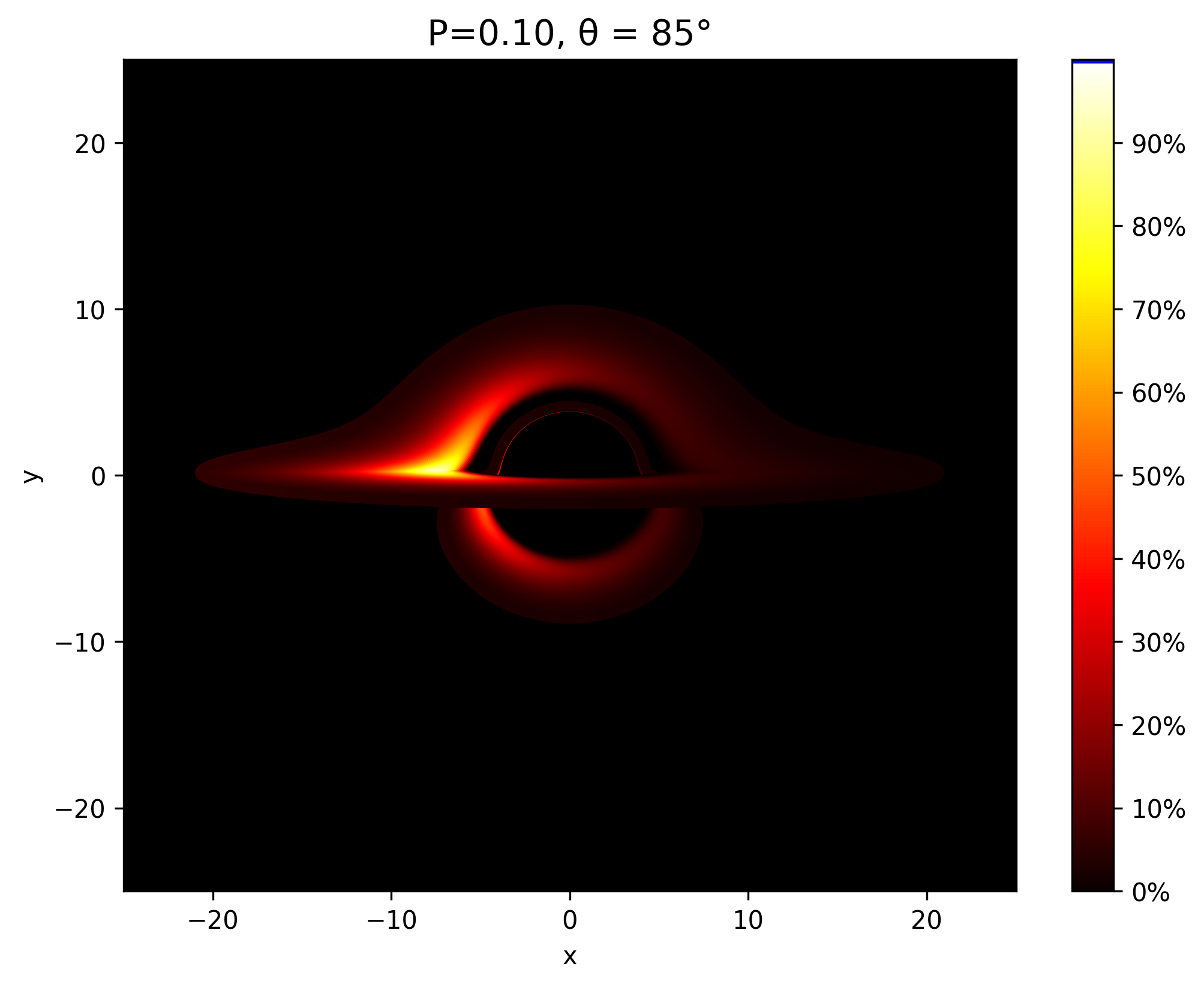}
  \end{tabular}
	\caption{\label{fig:fluxobs} 
The distribution of the observed flux $F_{\text{obs}}$ in the direct and secondary images of a self-dual BH in LQG is shown for inclination angles of $17^\circ$, $53^\circ$, and $85^\circ$. }
\end{figure*}

\begin{figure*}
\begin{tabular}{ccc}
  \includegraphics[scale=0.35]{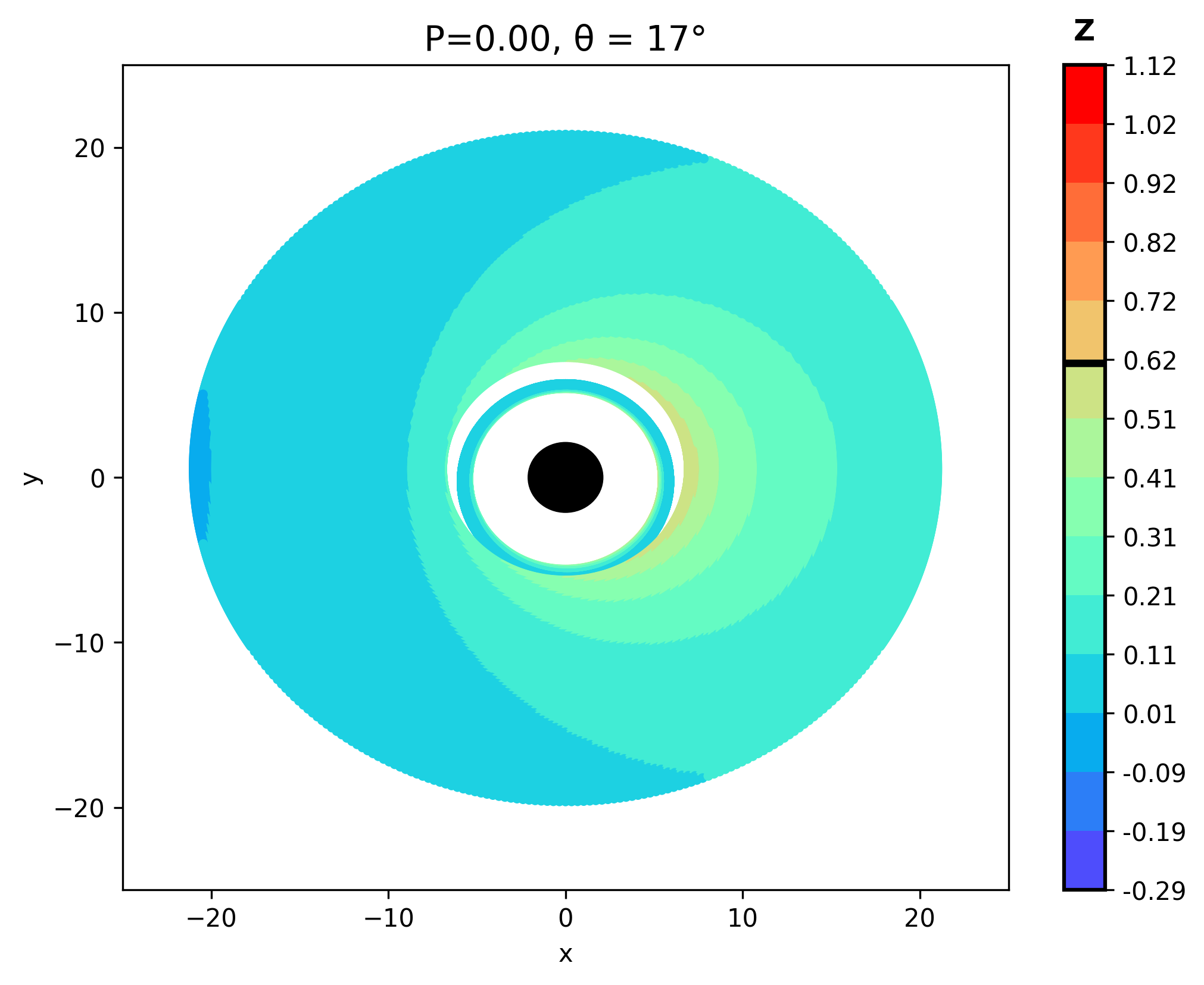}\hspace{-0.2cm}
  \includegraphics[scale=0.35]{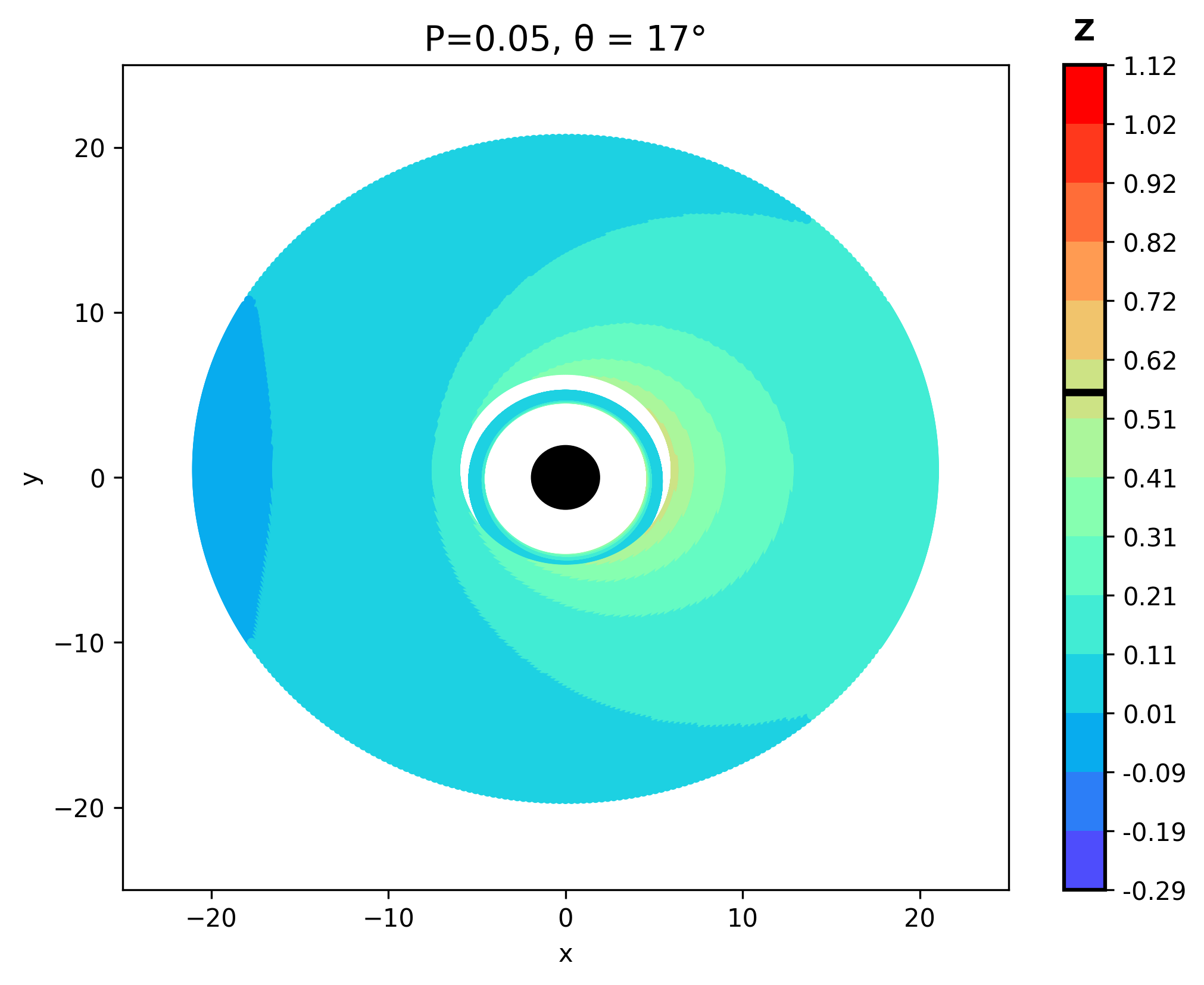}\hspace{-0.2cm}
  \includegraphics[scale=0.35]{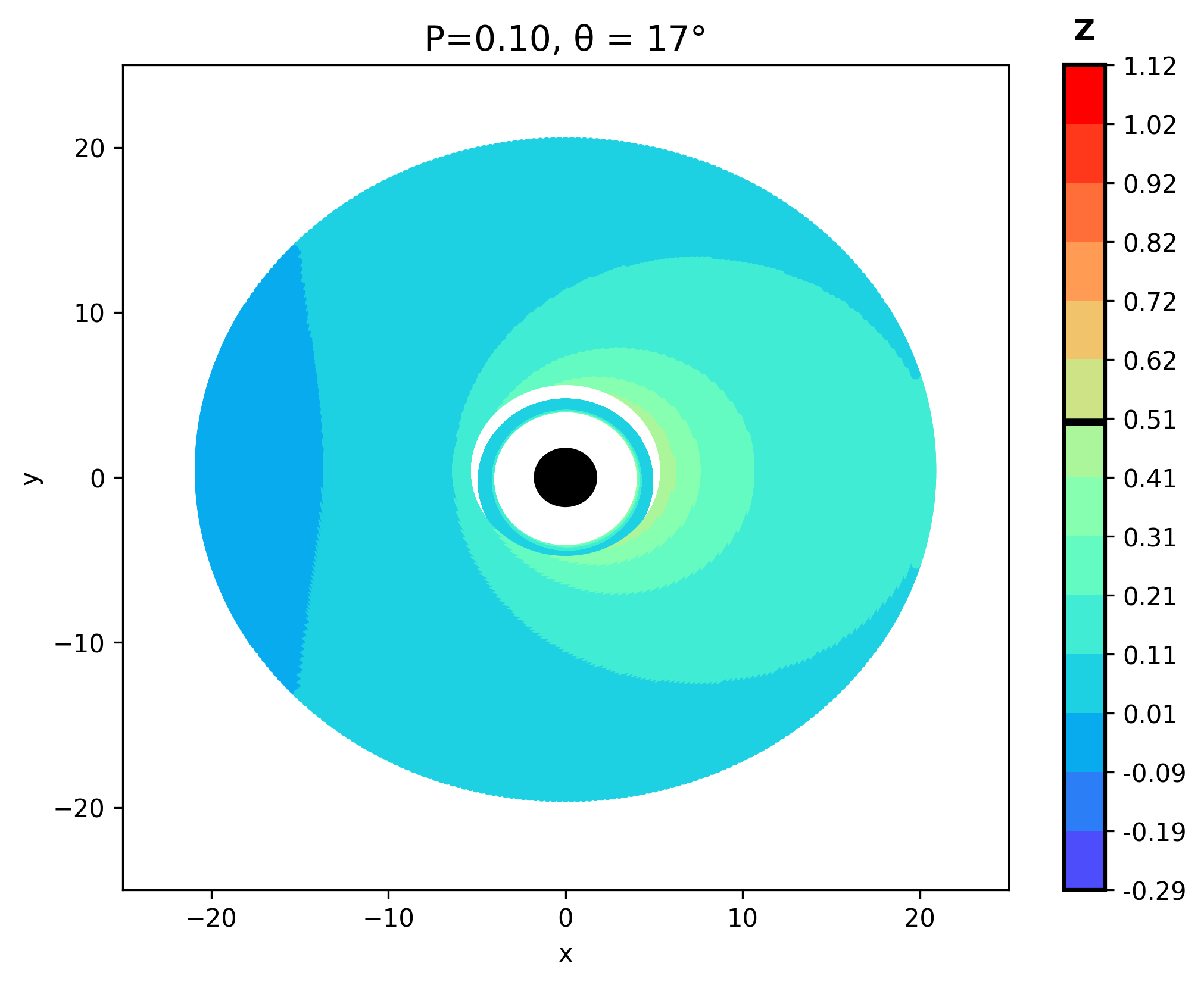}\\
  \includegraphics[scale=0.35]{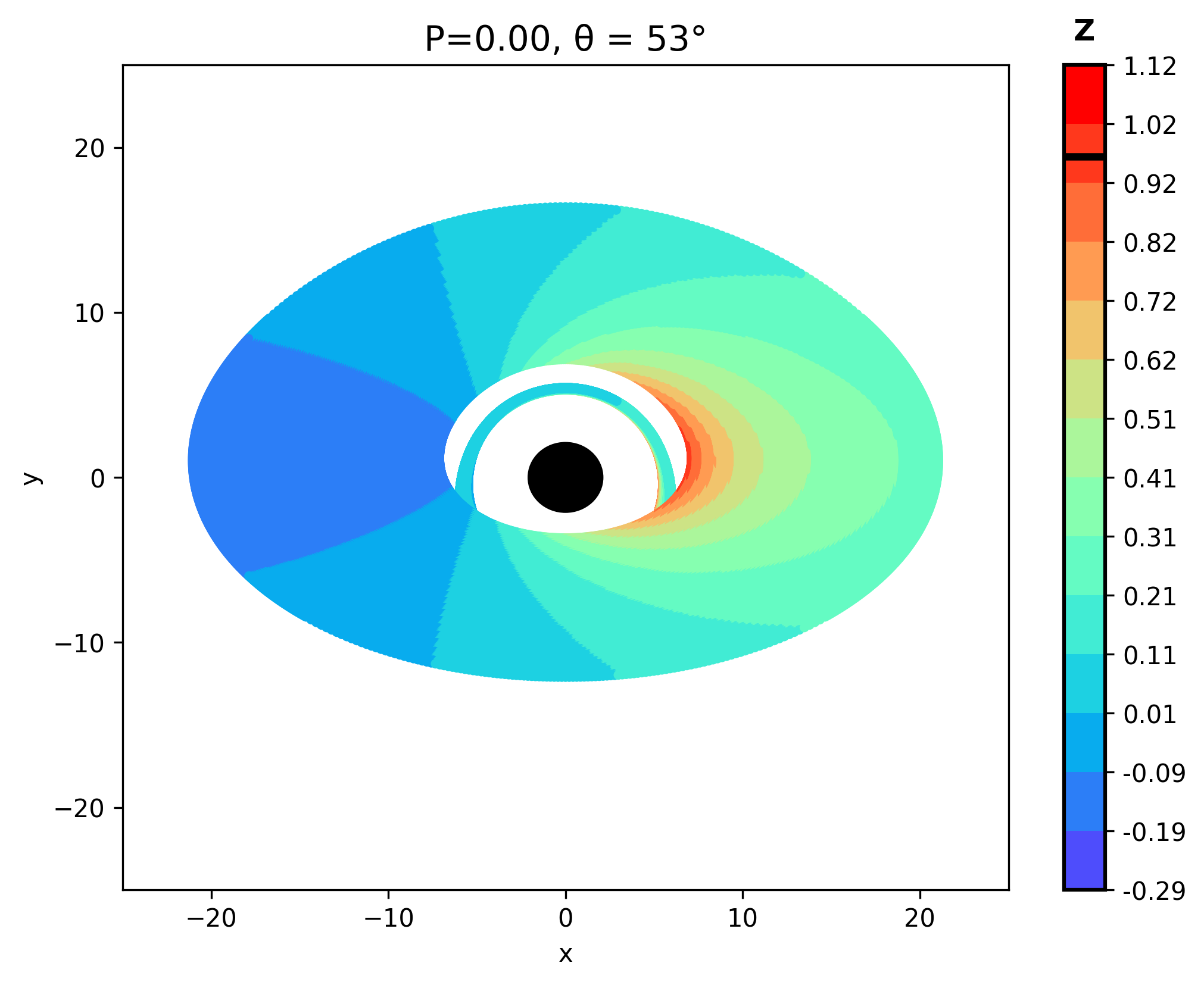}\hspace{-0.2cm}
  \includegraphics[scale=0.35]{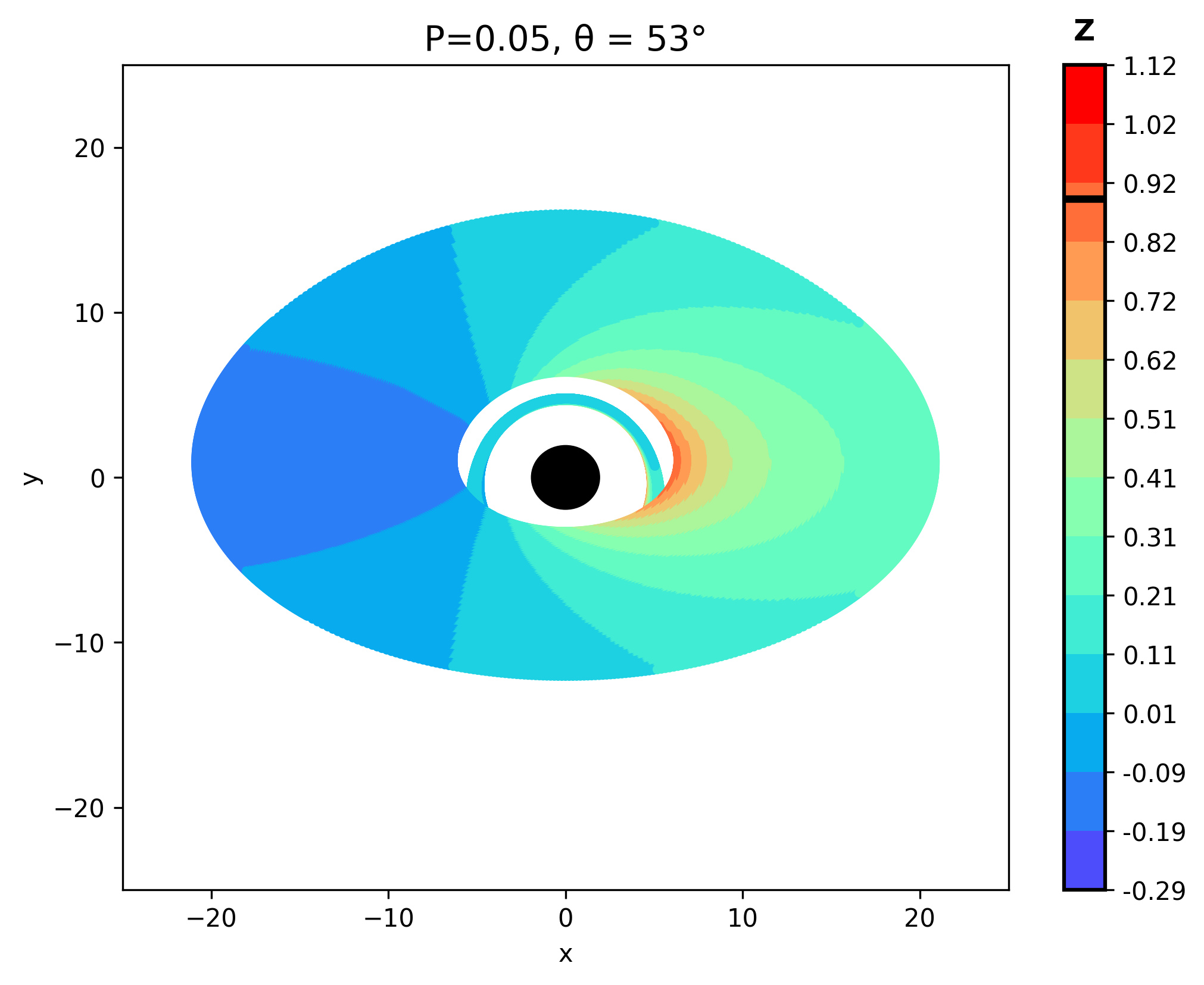}\hspace{-0.2cm}
  \includegraphics[scale=0.35]{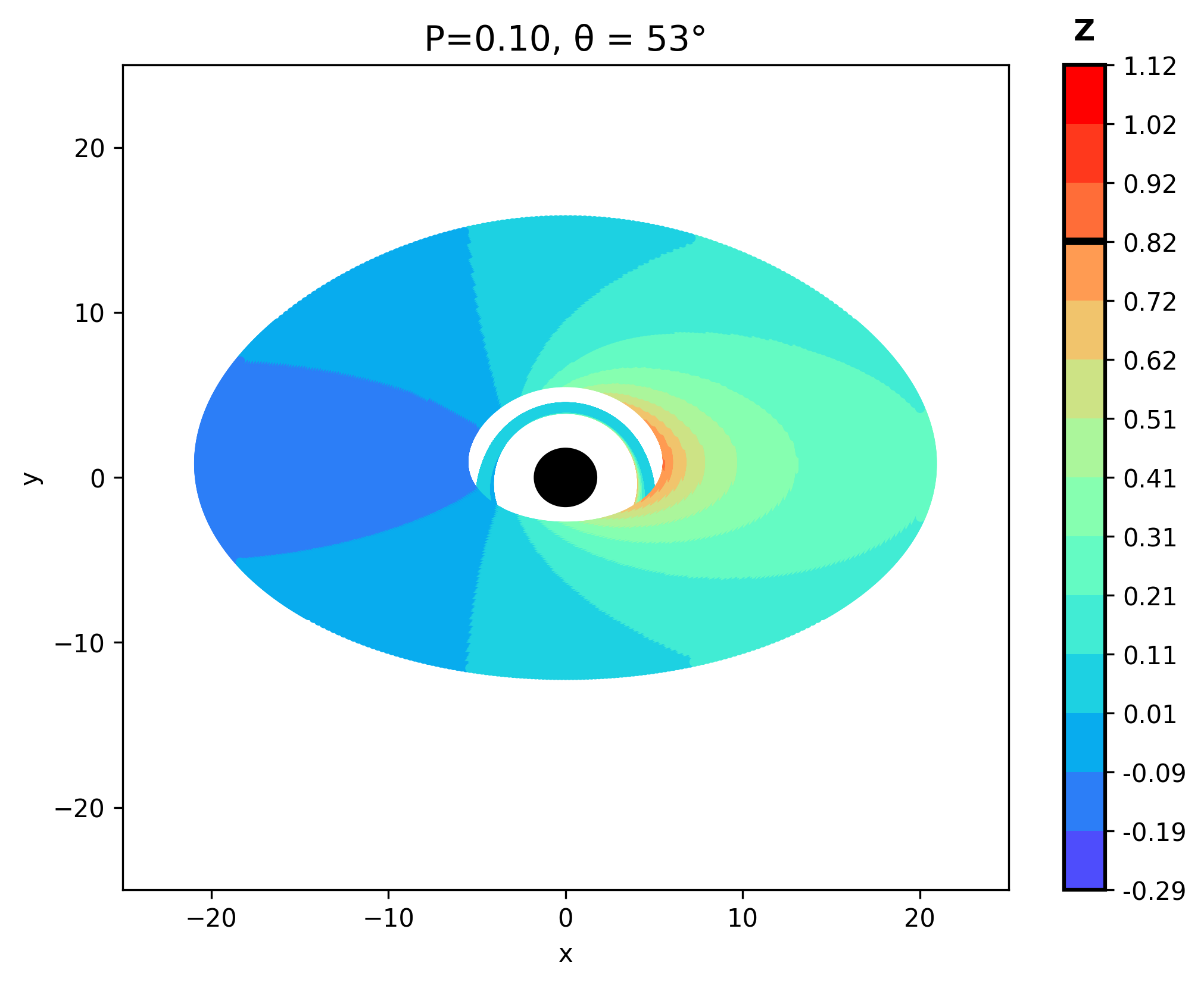}\\
  \includegraphics[scale=0.35]{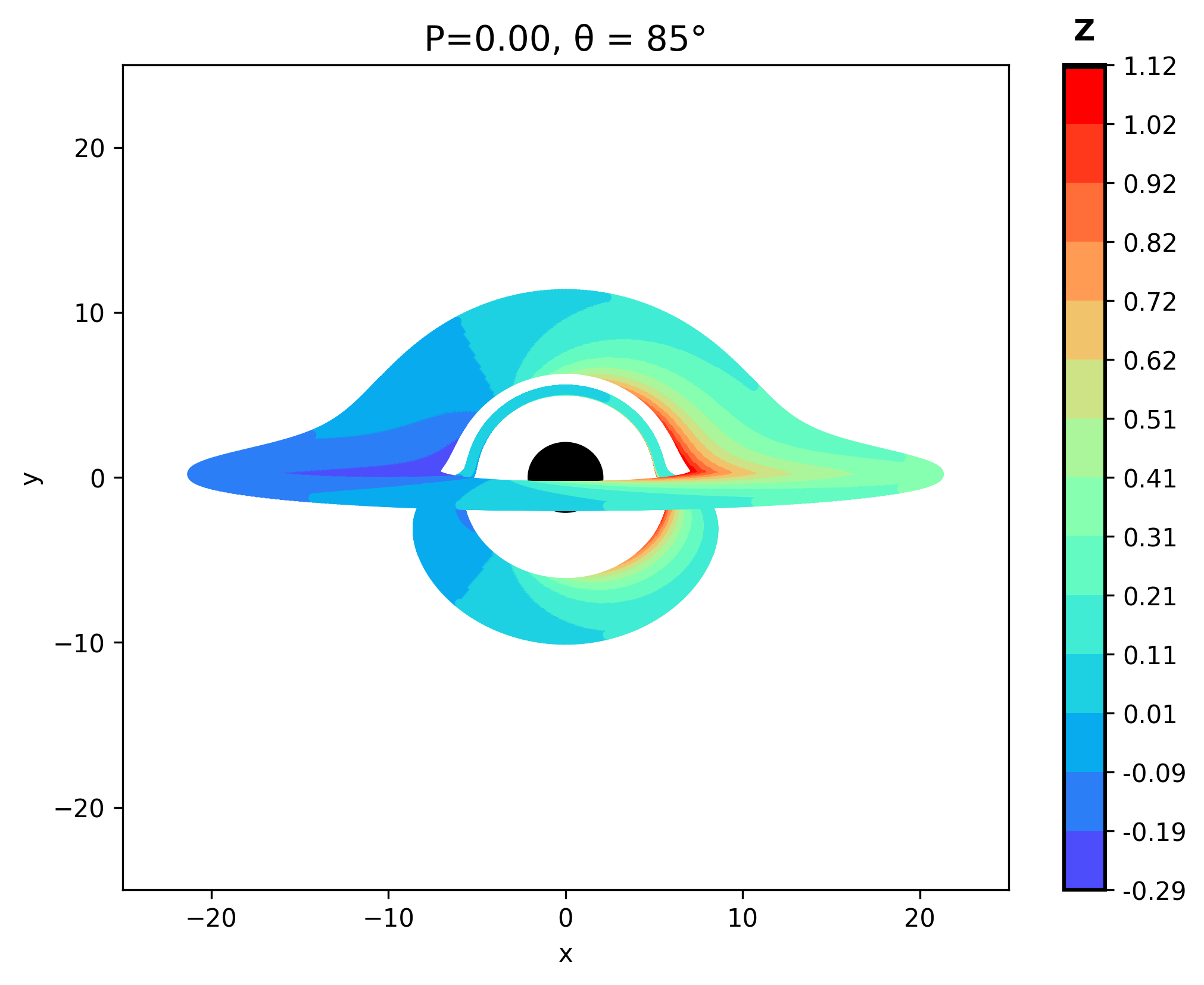}\hspace{-0.2cm}
  \includegraphics[scale=0.35]{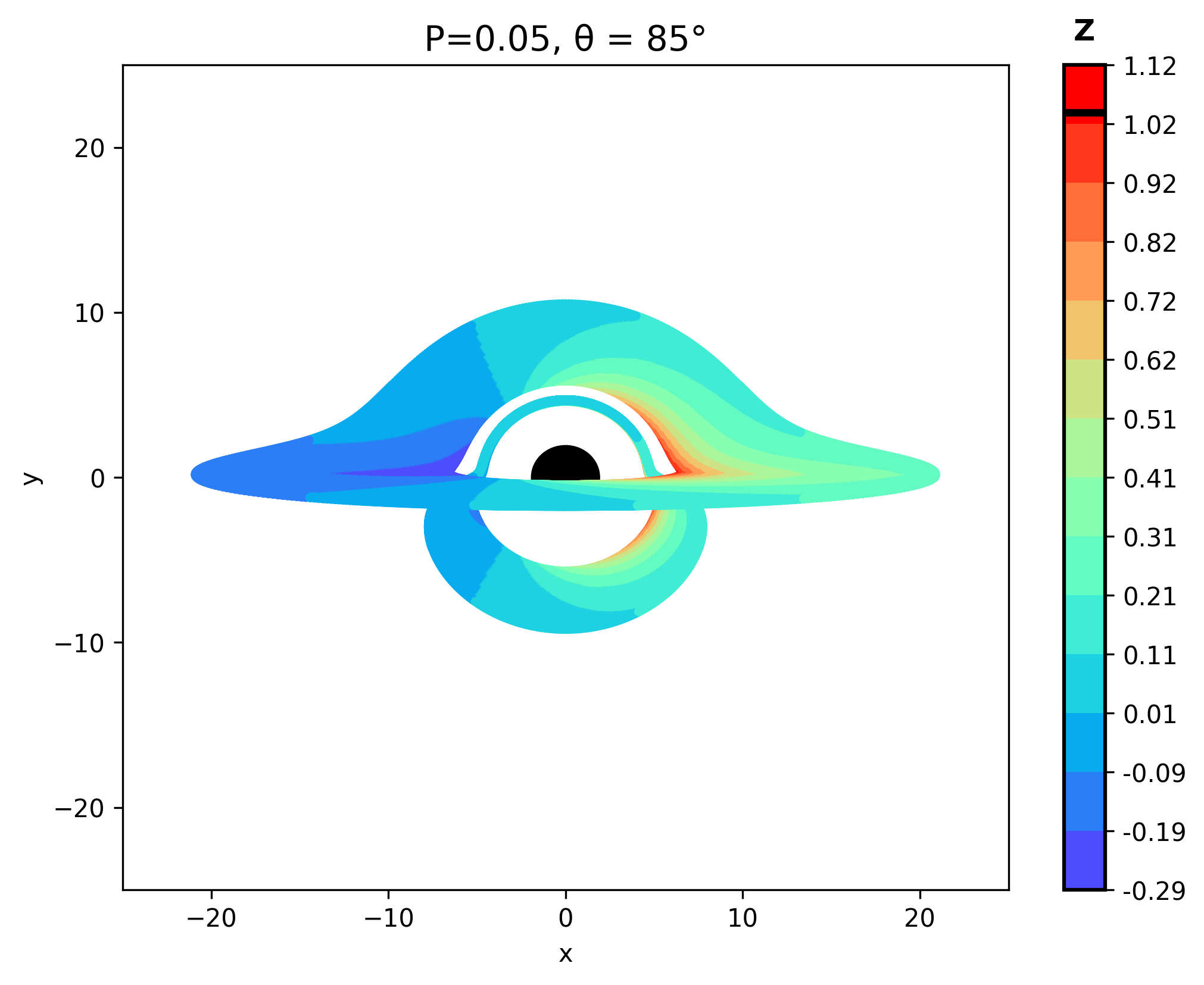}\hspace{-0.2cm}
  \includegraphics[scale=0.35]{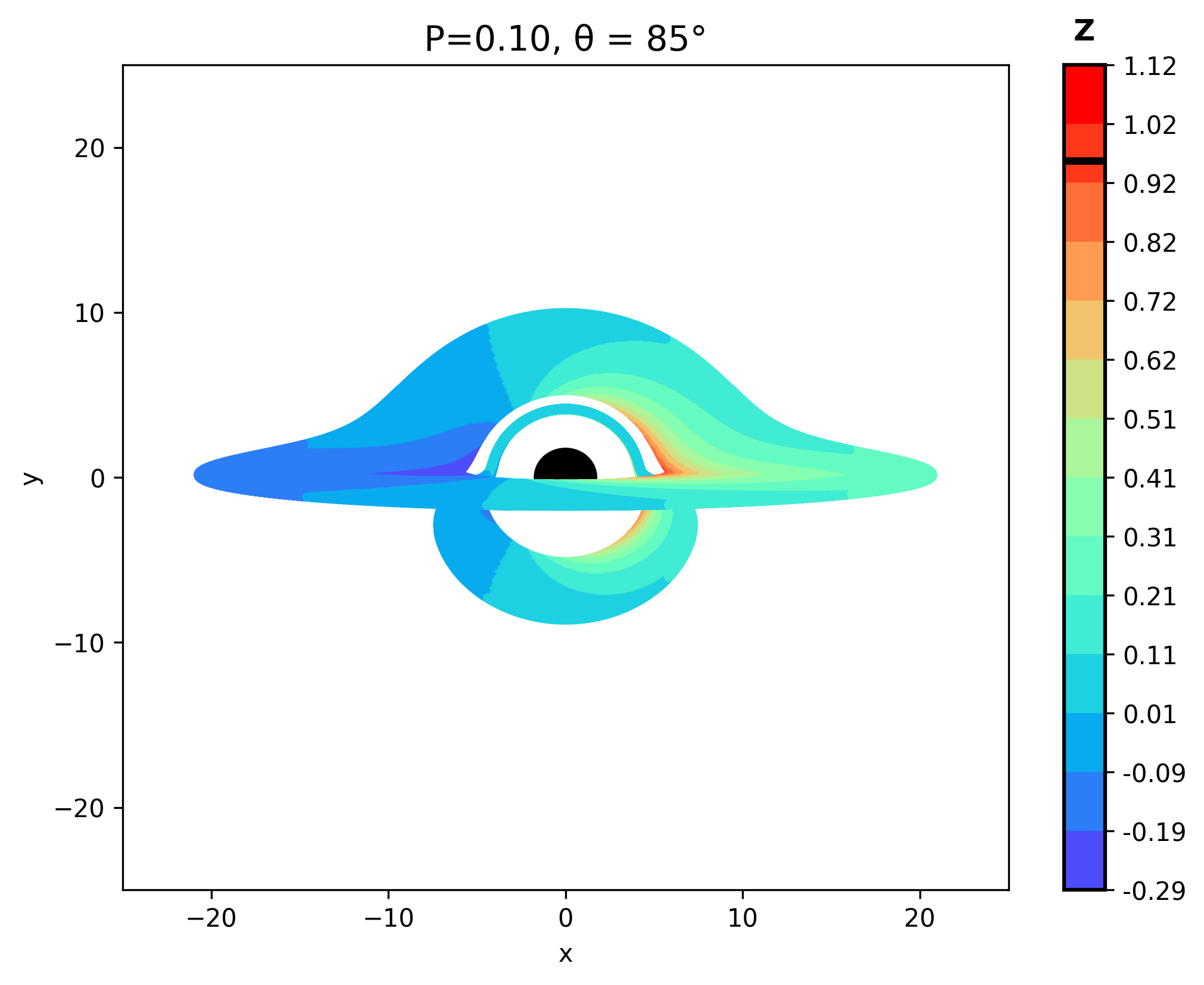}
  \end{tabular}
\caption{\label{fig:redeshiftdist}The distribution of the redshift factor $z$ in the direct and secondary images of a self-dual BH in LQG is presented for inclination angles of $17^\circ$, $53^\circ$, and $85^\circ$. }
    \end{figure*}

Applying the same method as in Fig.~\ref{fig:accrthin} to orbits from $R_{\text{isco}}$ to $R=25$, and assigning colors to the observed flux $F_{\text{obs}}$ and redshift factor $z$ corresponding to each $(b,\alpha)$, yields their distributions across the accretion disk of a self-dual BH in LQG (see Fig.~\ref{fig:fluxobs} and Fig.~\ref{fig:redeshiftdist}). Fig.~\ref{fig:fluxobs} shows the distribution of $F_{\text{obs}}$ for various inclination angles and values of $P$. The figure shows that a self-dual BH in LQG and a Schwarzschild BH ($P=0$) exhibit similar $F_{\text{obs}}$ distributions. The key distinction is that the self-dual BH appears smaller and brighter than the Schwarzschild case. If the observed energy flux $F_{\text{obs}}$ of the self-dual BH in LQG at $P = 0.1$ and $\theta = 85^\circ$ is normalized to 100\%, then the observed energy flux of the Schwarzschild BH at the same inclination is about 80\%, as seen in the bottom row of Fig.~\ref{fig:fluxobs}.

Fig.~\ref{fig:redeshiftdist} shows the redshift distribution of the accretion disk for different inclination angles. Due to the symmetry of gravitational redshift alone, we would expect a symmetrical redshift ($z>0$) distribution. However, the rotation of the disk introduces a Doppler contribution, which makes the distribution asymmetric. On the side rotating toward the observer, the redshift is reduced and may even shift into blueshift ($z<0$), whereas on the opposite side the redshift is amplified. The black line on the color bar indicates the maximum redshift for each disk. For 
$P=0.0$, $\theta=85^\circ$
, the values 
$z_{max}=1.12$
and $z_{min}=-0.29$ are obtained. These correspond to the most extreme values obtained among all cases. The viewing angle also affects the distribution: stronger redshift effects appear at higher inclinations. One can deduce that the redshift behaviors of the self-dual BH in LQG and the Schwarzschild BH ($P=0$) are similar, though slightly weaker for the self-dual BH case. Such differences could serve as potential observational signatures to distinguish BH geometries.
% \subsection{Observed flux and Redshift factor}

\section{Conclusion}
\label{Sec:conclusion}

From an astrophysical perspective, LQG BHs differ from their classical counterparts and provide a framework to probe quantum-gravitational signatures, and their background geometries remain one of the most intriguing challenges in astrophysics. Motivated by this, in this work, we considered a self-dual BH in LQG, examining its spacetime geometry and exploring observational features of accretion disks, including their direct and secondary images and the redshift and flux of emitted radiation as measured by distant observers at infinity. First, we began investigating the motion of massive and massless particles around the self-dual BH in LQG. From the study of timelike particle motion, we derived constraints on the quantum correction parameter $P$ using observational data from the perihelion shift of Mercury and the orbit of the S2 star around Sgr A$^{\star}$. These constraints were found to be $P \leq 0.000043$ for Mercury and $P \leq 0.067419$ for the S2 star, respectively.

For massless particle motion, photon trajectories around the self-dual BH in LQG were plotted and compared with those around a Schwarzschild BH case (see Figs.~\ref{fig:ray1} and \ref{fig:ray2}). Our results show that the critical impact parameter $b_c$ corresponding to the photon sphere is $b_c = 5.196$ for the Schwarzschild BH case and $b_c = 4.803$ for the self-dual BH ($P = 0.03$). Moreover, it was demonstrated that light is less deflected around the self-dual BH compared to its counterpart, indicating a weakening of the gravitational field with an increasing the quantum correction parameter $P$. This behavior was further supported by the analysis of photon trajectories in various cases, as summarized in Table~\ref{tab:nb}.    

We further examined the angular deflection for the formation
of the direct and higher-order images. It was shown that the higher-order images in the region $n\geq1$ shrink slightly faster than direct images as the quantum correction parameter $P$ increases for a fixed accretion disk radius. However, the higher-order images of the rays emitted from different radii remain nearly unchanged compared to the direct images $n = 0$ (see Fig.~\ref{fig:FBline1}). To explore accretion disk properties, we employed the Novikov–Thorne model for a thin accretion disk. Images of the disk at different inclination angles and for various values of $P$ revealed that the self-dual BH in LQG appears smaller than its Schwarzschild counterpart, consistent with the weakening of the gravitational field as $P$ increases (see Fig.~\ref{fig:accrthin}).

Finally, we examined the distribution of the observed energy flux $F_{\text{obs}}$ and the redshift factor $z$ on the accretion disk. The analysis of the observed energy flux distribution across the accretion disk shows that the self-dual BH appears brighter than the Schwarzschild BH. At an inclination angle of $85^\circ$, the observed flux of the Schwarzschild BH is approximately 80\% of that of the self-dual BH with $P=0.1$; see Figs.~\ref{fig:FluxFR} and \ref{fig:fluxobs}. Although the redshift distributions of the two BHs appear nearly identical, the Schwarzschild BH exhibits a larger redshift factor than the self-dual BH in LQG. At an inclination angle of $\theta = 85^\circ$, the redshift factor is $z \approx 1.12$ for the Schwarzschild case, compared to $z \approx 0.95$ for the self-dual BH with $P = 0.1$ (see Fig.~\ref{fig:redeshiftdist}).

Overall, our findings highlight the principal differences between the self-dual BH in LQG and the Schwarzschild BH, suggesting that such distinctions may provide new insights into the physical nature and accretion properties of self-dual BHs and their observable signatures in future observations.

\acknowledgments

The research is supported by the National Natural Science Foundation of China under Grant No. W2433018.

% \appendix
\bibliographystyle{apsrev4-1}  
\bibliography{Ref1}

\end{document}